\documentstyle[12pt,rotate,psfig]{article} 
\input{psfig}

\newcommand{\lapprox}{%
\mathrel{%
\setbox0=\hbox{$<$}
\raise0.6ex\copy0\kern-\wd0
\lower0.65ex\hbox{$\sim$}
}}
\newcommand{\gapprox}{%
\mathrel{%
\setbox0=\hbox{$>$}
\raise0.6ex\copy0\kern-\wd0
\lower0.65ex\hbox{$\sim$}
}}
\newcommand{\be}{\begin{equation}}
\newcommand{\ee}{\end{equation}}
\newcommand{\bea}{\begin{eqnarray}}
\newcommand{\eea}{\end{eqnarray}}
\renewcommand{\theequation}{\arabic{section}.\arabic{equation}}
\newcommand{\lbl}[1]{\label{eq:#1}}
\newcommand{ \rf}[1]{(\ref{eq:#1})}

\newcommand{\order}{{\cal O}}

\newcommand{\FF}{{\cal F}_{\pi^0\gamma^*\gamma^*}}

\def\thefootnote{\fnsymbol{footnote}}

\begin{document}
\pagestyle{empty}
\begin{flushright}
November 6, 2001 \\ 
Revised: February 15, 2002 \\ 
CPT-2001/P.4253 \\
\end{flushright}

\vspace*{0.5cm}
\begin{center}
{\Large {\bf Hadronic light-by-light corrections to the muon $g-2$:\\[0.2cm]
 The pion-pole contribution}}\\[0.8cm]
Marc Knecht\footnote{knecht@cpt.univ-mrs.fr} 
and Andreas Nyf\/feler\footnote{nyf\/feler@cpt.univ-mrs.fr} \\[0.3cm]
Centre de Physique Th\'{e}orique, CNRS-Luminy, Case 907\\ 
    F-13288 Marseille Cedex 9, France

\indent

\indent

\indent

{\bf Abstract}\\
\end{center}
\noindent
The correction to the muon anomalous magnetic moment from the
pion-pole contribution to the hadronic light-by-light scattering is
considered using a description of the $\pi^0 - \gamma^* - \gamma^*$
transition form factor based on the large-$N_C$ and short-distance
properties of QCD. The resulting two-loop integrals are treated by
first performing the angular integration analytically, using the
method of Gegenbauer polynomials, followed by a numerical evaluation
of the remaining two-dimensional integration over the moduli of the
Euclidean loop momenta.  The value obtained,
$a_{\mu}^{\mbox{\tiny{LbyL;$\pi^0$}}}= +5.8~(1.0) \times 10^{-10}$,
disagrees with other recent calculations.  In the case of the vector
meson dominance form factor, the result obtained by following the same
procedure reads $a_{\mu}^{\mbox{\tiny{LbyL;$\pi^0$}}}\vert_{VMD}=+5.6
\times 10^{-10}$, and differs only by its overall sign from the value
obtained by previous authors. The inclusion of the $\eta$ and
$\eta^\prime$ poles gives a total value
$a_{\mu}^{\mbox{\tiny{LbyL;PS}}}= +8.3~(1.2) \times 10^{-10}$ for the
three pseudoscalar states. This result substantially reduces the
difference between the experimental value of $a_{\mu}$ and its
theoretical counterpart in the standard model.

\vspace*{1cm} 

\newpage

\renewcommand{\thefootnote}{\arabic{footnote}}
\setcounter{footnote}{0}

\pagestyle{plain}
\setcounter{page}{1}

\section{ Introduction}
\label{sec:intro} 
\renewcommand{\theequation}{\arabic{section}.\arabic{equation}}
\setcounter{equation}{0}

A high-precision measurement of the muon anomalous magnetic moment 
is presently being performed at the Brookhaven National Laboratory by 
the Muon $(g-2)$ Collaboration \cite{BNL1}. The value released recently 
\cite{BNL2},
\be
a_{\mu^+}\,=\,11\,659\,202~(14)~(6)\times 10^{-10}\,,
\lbl{BNLexp}
\ee
already improves the precision by a factor of 8 as compared to the
previous determination at CERN \cite{cern77}. The final aim of the
experiment is to reach a precision of $\pm4\times 10^{-10}$, after
completion of the analysis including the full set of data collected
during recent years.

If only contributions of (multiflavored) QED are taken into account,
the theoretical accuracy is still much better than the present
experimental one \cite{KNO90,KinMar}.  However, hadronic effects,
which certainly cannot be ignored in the case of the muon $g-2$
factor, are unfortunately more difficult to estimate, and the
corresponding uncertainties substantially increase, and actually
completely dominate, the error of the total theoretical
value. Nevertheless, the value \rf{BNLexp} disagrees with various
theoretical calculations in a way that could become very significant
as the BNL E821 experiment further decreases its statistical
error. Seen from this perspective, it is not surprising that the
announcement of the value \rf{BNLexp} has triggered a wealth of
theoretical activity (a complete bibliography of the recent work in
this direction would represent a tedious task \cite{spires}). It is, 
however, our feeling that, rather than calling for hasty conclusions
concerning possible evidence for physics beyond the standard model,
the present situation first requires that the hadronic contributions
be scrutinized anew and with greater care.

\begin{figure}[t]
\centerline{\psfig{figure=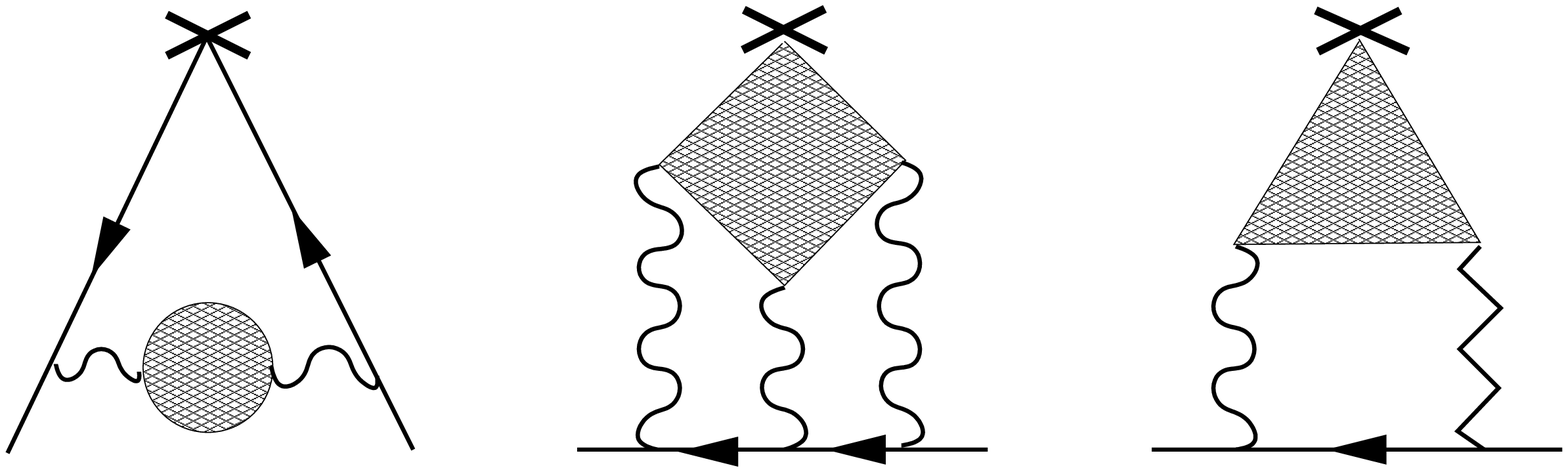,height=4.2cm,width=14cm
}}
\caption{The three topologies involving hadronic contributions to the  
anomalous magnetic moment of the muon: from left to right, vacuum
polarization insertion in the vertex, light-by-light scattering, and
two-loop electroweak contributions. The cross indicates an insertion
of the electromagnetic current, the shaded areas correspond to
hadronic subgraphs, while the lines exchanged in the rightmost graph
correspond to a photon and a neutral gauge boson.} \label{fig:fig1}
\end{figure}

The hadronic contributions to the anomalous magnetic moment of the
muon considered so far naturally fall into three categories: vacuum
polarization, light-by-light scattering, and higher-order electroweak
contributions, as represented in Fig.~\ref{fig:fig1}. Hadronic effects
in two-loop electroweak corrections are estimated in
Refs. \cite{PPdR95,CzKM95,KPPdR01}.  They are small, of the order of
the expected experimental error, and the associated theoretical
uncertainties, of the order of 10\%, can be brought under safe
control~\cite{KPPdR01}.  Hadronic vacuum polarization effects benefit
from the fact that they can be related to the total cross section for
$e^+e^-\to {\mbox{hadrons}}$ \cite{early_hadvacpol}.  However,
existing data come from different sources, and do not always meet the
required accuracy, so that they are supplemented with theoretical
input. There exist therefore different estimates for the hadronic
vacuum polarization, which either strengthen the difference between
theory and the experimental value \rf{BNLexp}, or make it appear only
marginal. Our purpose here is not to intervene in this ongoing debate
\cite{davier,narison01,jegerlehner01,yndurain01,cvetic01,prades01,cirigliano},
but rather to have a closer look at the remaining item on the list,
namely, the hadronic effects in the so-called light-by-light scattering
contribution. There exist several estimates of the latter
\cite{calmet76,KNO_85,HKS_95_96,HK_98,BPP}. The latest to date,
obtained by two different groups (see also \cite{bartos01}), 
give values that are consistent
within the quoted errors,
\bea
a_{\mu}^{\mbox{\tiny{LbyL; had}}} &=& -7.9~(1.5)\times 10^{-10}
\quad\cite{HKS_95_96,HK_98}\,,
\lbl{th1}
\\
a_{\mu}^{\mbox{\tiny{LbyL; had}}} &=& -9.2~(3.2)\times 10^{-10}
\quad\cite{BPP}\,.
\lbl{th2}
\eea
On the other hand, both groups have changed their value of
$a_{\mu}^{\mbox{\tiny{LbyL; had}}}$, {\it and even its sign}, at some
stage, and for reasons that are not very easy to follow -- partly
because both analyses rapidly leave the realm of analytical work and
resort to numerical techniques.  This aspect of the present
theoretical situation provided the motivation to improve from the
analytical side the study of the hadronic light-by-light
contributions.

A complete discussion of the hadronic light-by-light contributions
involves the full rank-four hadronic vacuum polarization tensor
$\Pi_{\mu\nu\lambda\rho}(q_1,q_2,q_3)$, which is a rather involved
object. On the other hand, the values \rf{th1} and \rf{th2} are both
dominated by a well-identified component, $a_{\mu}^{\mbox{\tiny{LbyL;
had}}}=a_{\mu}^{\mbox{\tiny{LbyL;}}\pi^0}+\cdots$,
\bea
a_{\mu}^{\mbox{\tiny{LbyL;}}\pi^0} &=& -5.7~(0.3)\times 10^{-10}
\quad\cite{HK_98}\,,
\lbl{th1pi0}
\\
a_{\mu}^{\mbox{\tiny{LbyL;}}\pi^0} &=& -5.9~(0.9)\times 10^{-10}
\quad\cite{BPP}\,,
\lbl{th2pi0}
\eea
arising from the one-particle reducible pion-exchange pieces of
$\Pi_{\mu\nu\lambda\rho}(q_1,q_2,q_3)$. In the present paper, we focus
on the latter, leaving for future work a complete discussion of
$\Pi_{\mu\nu\lambda\rho}(q_1,q_2,q_3)$ within a theoretical framework
close to the one adopted here (and discussed below) for the pion-pole
contribution.

\begin{figure}[t]
\centerline{\psfig{figure=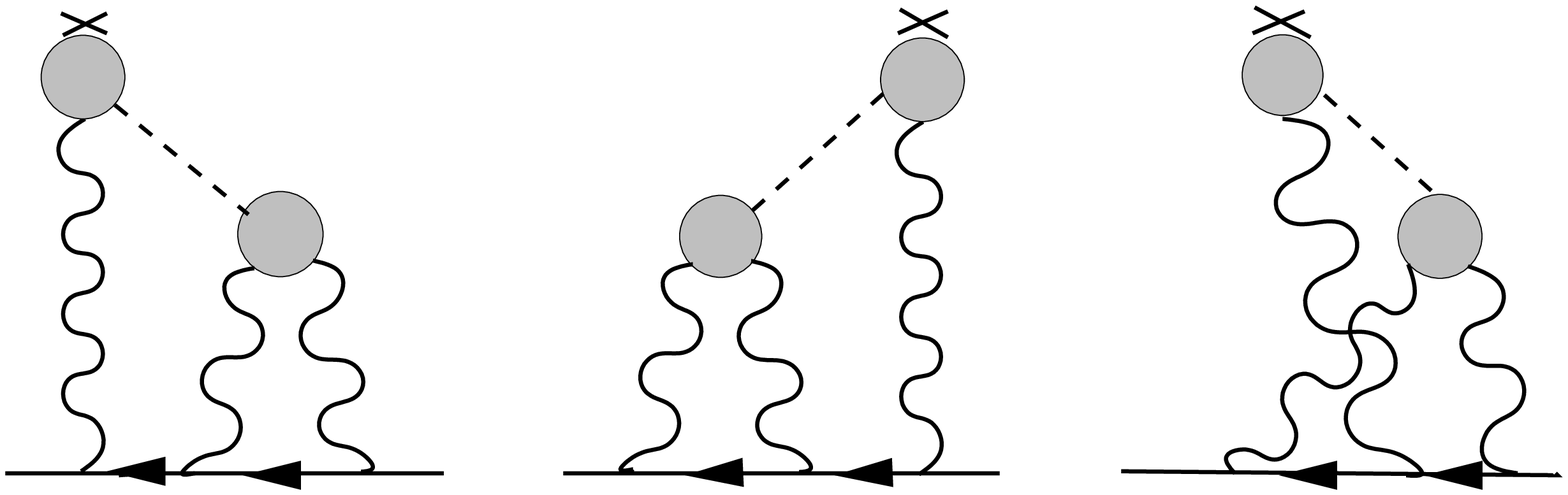,height=4.6cm,width=16cm
}}
\caption{The pion-pole contributions to light-by-light scattering. The shaded 
blobs represent the form factor $\FF$. The first and second graphs
give rise to identical contributions, involving the function
$T_1(q_1,q_2;p)$ in Eq. \protect\rf{a_pion_2}, whereas the third graph
gives the contribution involving $T_2(q_1,q_2;p)$. }
\label{fig:fig2}
\end{figure}

The main ingredient in the determination of the pion-exchange graphs
depicted in Fig.~\ref{fig:fig2} is the double off-shell
pion-photon-photon transition form factor ${\cal
F}_{\pi^0\gamma^*\gamma^*}(q_1^2,q_2^2)$. No sufficiently complete
experimental information on this quantity is available at the time
being, so that even the one-pion-exchange component of
$\Pi_{\mu\nu\lambda\rho}(q_1,q_2,q_3)$ is beyond reach from an
experimental point of view. Resort to theory is therefore
unavoidable. Now, from the QCD perspective, the theoretical knowledge
of $\FF$ is sparse. There exists a well-defined limit of QCD where the
situation improves somewhat, namely, the limit of infinite number of
colors $N_C$ \cite{tHooft74,witten79}.  As $N_C$ becomes large, the
spectrum of QCD reduces, in each channel with given quantum numbers,
to an infinite tower of zero-width resonances. As a consequence, the
analytic structure of the form factor $\FF(q_1^2,q_2^2)$ becomes
simpler: it consists of a succession of simple poles, due to the
contributions of zero-width $J^{PC}=1^{--}$ states (e.g., the $\rho$
meson and its radial excitations) in each channel. Dealing with an
infinite number of resonances remains cumbersome, and to a large
extent illusory, since the characteristics of these states (masses,
couplings, etc.) are in general not known. It has, however, been shown
in several instances (see, e.g., the review \cite{derafael01} and
references therein) that keeping, in each channel, only a finite
number of resonances, supplemented with information on the QCD
short-distance properties coming from the operator product expansion
\cite{wilson69,SVZ}, already gives a good description of quantities
like form factors or correlation functions in the Euclidean region,
especially when they occur in weighted integrals over the whole range
of momenta.  In particular, the form factor $\FF$ has recently been
studied \cite{paper_VAP} from the point of view of this lowest meson
dominance (LMD) or minimal hadronic {\it Ansatz} (MHA) approximation to
large-$N_C$ QCD.  Since the analyses carried out in
Refs. \cite{HK_98,BPP} rely on models of $\FF$, vector meson dominance
(VMD) or extended Nambu--Jona-Lasinio (ENJL) (see \cite{BPP} and
references therein), that do not reproduce the correct QCD
short-distance properties, a second motivation for the present study
was to compare the results \rf{th1pi0} and \rf{th2pi0} with those
derived from a representation of $\FF$ that complies with these QCD
constraints.  Finally, let us mention for completeness that the
pion-pole contribution we are interested in corresponds to the
lowest-mass part of $\Pi_{\mu\nu\lambda\rho}(q_1,q_2,q_3)$ that is
leading in the large-$N_C$ limit \cite{derafael94}, which might
provide an explanation as to why it happens to constitute the dominant
fraction of the light-by-light scattering correction to $a_{\mu}$.

The remaining material of this paper is organized as
follows. Section~\ref{sec:def} recalls a few definitions that are
relevant for the light-by-light contribution to $a_\mu$, and also
serves the purpose of introducing our notation. The expression for the
two-loop integral which gives the pion-pole contribution
$a_{\mu}^{\mbox{\tiny{LbyL;$\pi^0$}}}$ to the anomalous magnetic
moment of the muon in terms of $\FF$ is derived in
Sec.~\ref{sec:pionpole}.  In Sec.~\ref{sec:formfactors} we discuss a
generic class of form factors to which those inspired from large-$N_C$
QCD and considered here belong, but that also covers other cases, like
the constant form factor, given by the Wess-Zumino-Witten term
\cite{WZW}, or the vector meson dominance form factor. The method of
Gegenbauer polynomials is presented in Sec.~\ref{sec:angular} and used
in order to perform the angular integrations.  This then leads to a
two-dimensional integral representation for
$a_{\mu}^{\mbox{\tiny{LbyL;$\pi^0$}}}$ in terms of $\FF$ and of
several weight functions
(Sec.~\ref{sec:twodimrepresentation}). Numerical results for
$a_{\mu}^{\mbox{\tiny{LbyL;$\pi^0$}}}$ are presented in
Sec.~\ref{sec:numerics}, where the contributions from the $\eta$ and
$\eta^\prime$ poles are also briefly discussed. A discussion and
conclusions make up the content of Sec.~\ref{sec:conclusions}. For the
reader's convenience, the properties of the form factor $\FF$ that
form the basis of the representations we use here have been gathered
in an Appendix.

\section{Definitions}
\label{sec:def} 
\renewcommand{\theequation}{\arabic{section}.\arabic{equation}}
\setcounter{equation}{0}

The interaction of the photon field with the standard model fermionic
degrees of freedom is described by the Lagrangian ($e$ denotes the
electric charge of the electron)
\be
{\cal L}_{\mbox{\tiny{int}}}(x)\,=\,-eA_{\rho}(x)\,\bigg[
\sum_{\ell} 
({\overline\psi}_{\ell}\gamma^{\rho}\psi_{\ell})(x)\,-
\,\sum_{q}e_q({\bar q}\gamma^{\rho}q)(x)
\bigg]
\,.
\ee
The first sum runs over the charged lepton flavors $\ell = e^-\! \!
,\mu^-\!\! ,\tau^-$, while the second sum runs over the quark flavors,
with $e_q$ standing for the corresponding electric charges expressed
in units of $\vert e\vert$.  Our interest hereafter is restricted to
the contributions of the light quarks $q=u,d,s$.  The muon
gyromagnetic ratio is obtained from the muon proper vertex function
$\Gamma_{\rho}(p\,',p)$, defined as ($p^2=p\,'{}^2=m^2$, $m$ denotes
the muon mass)
\be
(-ie){\bar{\mbox{u}}}(p\,')\Gamma_{\rho}(p\,',p)\mbox{u}(p) \,=\,
\langle \mu^-(p\,')\vert (-ie)\sum_{\ell}
({\overline\psi}_{\ell}\gamma_{\rho}\psi_{\ell})(0)\,+
\,(ie)\sum_{q}e_q({\bar q}\gamma_{\rho}q)(0)
\ \vert \mu^-(p)\rangle
\,,
\ee
at vanishing momentum transfer.

The contribution to $\Gamma_{\rho}(p\,',p)$ of relevance here is the
matrix element, at lowest nonvanishing order in the fine structure
constant $\alpha = e^2/(4\pi)$, of the light quark electromagnetic
current
\be
j_{\rho}(x) \,=\, \frac{2}{3}({\bar u}\gamma_{\rho}u)(x)\,-
\,\frac{1}{3}({\bar d}\gamma_{\rho} d)(x)
\,-\,\frac{1}{3}({\bar s}\gamma_{\rho} s)(x)
\lbl{current}
\ee
between $\mu^-$ states,
\bea
(-ie){\bar{\mbox{u}}}(p\,'){\widehat\Gamma}_{\rho}(p\,',p)\mbox{u}(p)
&\equiv & \langle \mu^-(p\,')\vert (ie)j_{\rho}(0) \vert \mu^-(p)\rangle
\nonumber\\
&=&
\int\frac{d^4q_1}{(2\pi)^4}\int\frac{d^4q_2}{(2\pi)^4}\,
\frac{(-i)^3}{q_1^2\,q_2^2\,(q_1+q_2-k)^2}\,
\nonumber\\
&
\times &\!\!\!\!\!
\frac{i}{(p\,'-q_1)^2-m^2}\,
\frac{i}{(p\,'-q_1-q_2)^2-m^2}
\nonumber\\
&
\times &\!\!\!\!\!
(-ie)^3{\bar{\mbox{u}}}(p\,')
\gamma^{\mu}(\not\! p\,'- \not\!q_1+m)
\gamma^{\nu}(\not\! p\,'- \not\! q_1-\not\! q_2+m)
\gamma^{\lambda}\mbox{u}(p)
\nonumber\\
&
\times &\!\!\!\!\!
(ie)^4\Pi_{\mu\nu\lambda\rho}(q_1,q_2,k-q_1-q_2)\,,
\eea
with $k_{\mu}=(p\,' - p)_{\mu}$ and 
\bea
\Pi_{\mu\nu\lambda\rho}(q_1,q_2,q_3) &=&
\int d^4x_1\int d^4x_2\int d^4x_3
\,e^{i(q_1\cdot x_1 + q_2\cdot x_2 + q_3\cdot x_3)}\,
\nonumber\\
&&\quad
\times
\langle\,\Omega\,\vert\,\mbox{T}
\{j_{\mu}(x_1)j_{\nu}(x_2)j_{\lambda}(x_3)j_{\rho}(0)\}  
\,\vert\,\Omega\,\rangle
\eea
the fourth-rank light quark hadronic vacuum polarization tensor,
$\vert\,\Omega\,\rangle$ denoting the QCD vacuum.

Since the flavor diagonal current $j_{\mu}(x)$ is conserved, the tensor 
$\Pi_{\mu\nu\lambda\rho}(q_1,q_2,q_3)$ satisfies the Ward identities
\be
\{q_1^{\mu};q_2^{\nu};q_3^{\lambda};(q_1+q_2+q_3)^{\rho}\}
\Pi_{\mu\nu\lambda\rho}(q_1,q_2,q_3)\,=\,0
\,.
\ee
This entails that $\Pi_{\mu\nu\lambda\rho}(q_1,q_2,k-q_1-q_2) = -
k^\sigma (\partial / \partial k^\rho)
\Pi_{\mu\nu\lambda\sigma}(q_1,q_2,k-q_1-q_2)$ and\footnote{We use
the following conventions for 
Dirac's $\gamma$ matrices:
$\{\gamma_{\mu},\gamma_{\nu}\}=2\eta_{\mu\nu}$, with $\eta_{\mu\nu}$
the flat Minkowski space metric of signature $(+~-~-~-)$,
$\sigma_{\mu\nu}=(i/2)[\gamma_{\mu},\gamma_{\nu}]$,
$\gamma_5=i\gamma^0\gamma^1\gamma^2\gamma^3$, whereas the totally
antisymmetric tensor $\varepsilon_{\mu\nu\rho\sigma}$ is chosen such
that $\varepsilon_{0123}=+1$. }
\be
{\bar{\mbox{u}}}(p\,'){\widehat\Gamma}_{\rho}(p\,',p)\mbox{u}(p)\,=\,
{\bar{\mbox{u}}}(p\,')\bigg[ \gamma_{\rho}{\widehat F}_1(k^2)\,+\,
\frac{i}{2m}\,\sigma_{\rho\tau}k^{\tau}{\widehat F}_2(k^2)\bigg]\mbox{u}(p)
\,,
\ee
as well as
${\widehat\Gamma}_{\rho}(p\,',p) =
k^{\sigma}{\widehat\Gamma}_{\rho\sigma}(p\,',p)$ with
\bea
{\bar{\mbox{u}}}(p\,'){\widehat\Gamma}_{\rho\sigma}(p\,',p)\mbox{u}(p) 
&=&
-ie^6\,\int\frac{d^4q_1}{(2\pi)^4}\int\frac{d^4q_2}{(2\pi)^4}\,
\frac{1}{q_1^2\,q_2^2\,(q_1+q_2-k)^2}\,
\nonumber\\
&&\quad
\times
\frac{1}{(p\,'-q_1)^2-m^2}\,
\frac{1}{(p\,'-q_1-q_2)^2-m^2}
\nonumber\\
&&\quad
\times
{\bar{\mbox{u}}}(p\,')
\gamma^{\mu}(\not\! p\,'- \not\!q_1+m)
\gamma^{\nu}(\not\! p\,'- \not\! q_1-\not\! q_2+m)
\gamma^{\lambda}\mbox{u}(p)
\nonumber\\
&&\quad
\times
\frac{\partial}{\partial k^{\rho}}\,
\Pi_{\mu\nu\lambda\sigma}(q_1,q_2,k-q_1-q_2)\, . 
\lbl{Gamma2}
\eea

Following Ref. \cite{ABDK} and using the property
$k^{\rho}k^{\sigma}{\bar{\mbox{u}}}(p\,')
{\widehat\Gamma}_{\rho\sigma}(p\,',p)\mbox{u}(p)=0$, one deduces that
${\widehat F}_1(0)=0$ and that the hadronic light-by-light
contribution to the muon anomalous magnetic moment is equal to
\be
{\widehat F}_2(0)\,=\,
\frac{1}{48m}\,
{\mbox{tr}}\left\{(\not\! p + m)[\gamma^{\rho},\gamma^{\sigma}](\not\! p + m)
{\widehat\Gamma}_{\rho\sigma}(p,p)\right\}
\,.
\lbl{F2trace}
\ee
Equivalently, one may project out ${\widehat F}_2(k^2)$ from  
${\widehat\Gamma}_{\rho}(p\,',p)$,
\be
{\widehat F}_2(k^2)\,=\,
{\mbox{tr}}\left\{(\not\! p + m)\Lambda^{(2)}_{\rho}(p\,',p) (\not\! p\,'+ m)
{\widehat\Gamma}^{\rho}(p\,',p)\right\}\,,
\lbl{F2proj}
\ee
with the help of the projector \cite{BS67}
\be
\Lambda^{(2)}_{\rho}(p\,',p) 
\,=\,\frac{m^2}{k^2(4m^2-k^2)}\,\bigg[\gamma_{\rho}\,+\,
\frac{k^2+2m^2}{m(k^2-4m^2)}\,(p\,'+p)_{\rho}\bigg]\,.
\lbl{proj1}
\ee
One then uses ${\widehat\Gamma}^{\rho}(p\,',p) = k_\sigma
{\widehat\Gamma}^{\rho\sigma}(p\,',p)$ and employs the identity
(for $p^2=p\,'{}^2=m^2$) 
\be
(\not\! p + m)\, \gamma_{\rho} \, (\not\! p\,'+ m) = (\not\! p + m)\,
\left[ {1 \over 2m} (p + p\,')_\rho + {i \over 2m}
\sigma_{\rho\kappa} (p - p\,')^\kappa \right] \, (\not\! p\,'+ m)   
\ee  
to simplify $\Lambda^{(2)}_{\rho}(p\,',p)$.  Averaging over the
directions of the four-vector $k_{\mu}$ with the constraints $k \cdot
p = - k^2/2$, $k \cdot p\,' = k^2/2$ and keeping the leading terms in
$k$ in the resulting expression then gives back
Eq.~\rf{F2trace} \cite{BR,RRL}.

\section{Pion-pole contribution}
\label{sec:pionpole} 
\renewcommand{\theequation}{\arabic{section}.\arabic{equation}}
\setcounter{equation}{0}

From now on, we concentrate on the contributions to 
$\Pi_{\mu\nu\lambda\rho}(q_1,q_2,q_3)$ arising from single 
neutral pion exchanges (see Fig.~\ref{fig:fig2}), which read
\bea
\Pi_{\mu\nu\lambda\rho}^{(\pi^0)}(q_1,q_2,q_3) & = & i
\,{\FF(q_1^2, q_2^2) \ \FF(q_3^2, (q_1+q_2+q_3)^2) \over (q_1+q_2)^2 - 
M_{\pi}^2} \ \varepsilon_{\mu\nu\alpha\beta} \, q_1^\alpha q_2^\beta \ 
\varepsilon_{\lambda\rho\sigma\tau} \, q_3^\sigma (q_1 + q_2)^\tau
\nonumber \\ 
& &
\!\! \,+ i\, {\FF(q_1^2, (q_1 + q_2 + q_3)^2) \ \FF(q_2^2, q_3^2) \over
(q_2+q_3)^2 - M_{\pi}^2} \ \varepsilon_{\mu\rho\alpha\beta} \,
q_1^\alpha (q_2 + q_3)^\beta \ \varepsilon_{\nu\lambda\sigma\tau} \,
q_2^\sigma q_3^\tau \nonumber \\
&&
\!\! +\, i\, {\FF(q_1^2, q_3^2) \ \FF(q_2^2, (q_1+q_2+q_3)^2) \over
(q_1+q_3)^2 - M_{\pi}^2} \ \varepsilon_{\mu\lambda\alpha\beta} \,
q_1^\alpha q_3^\beta \  \varepsilon_{\nu\rho\sigma\tau} \, q_2^\sigma
(q_1 + q_3)^\tau . \nonumber \\ 
& & 
\eea
The form factor $\FF(q_1^2, q_2^2)$ is defined as (using the same
convention as in Ref.~\cite{paper_VAP}) 
\be
i \int d^4 x e^{i q \cdot x} \langle\,\Omega | T \{ j_\mu(x) j_\nu(0)
\} | \pi^0(p) \rangle \,=\, \varepsilon_{\mu\nu\alpha\beta} \, q^\alpha
p^\beta \, \FF(q^2,(p-q)^2)  
\ee 
with $\FF(q_1^2,q_2^2) = \FF(q_2^2,q_1^2)$.  For the computation of
the corresponding value of
$a_{\mu}^{\mbox{\tiny{LbyL;$\pi^0$}}}\equiv{\widehat
F}_2(0)\vert_{\mbox{\tiny{pion pole}}}$, we need 
\bea
\frac{\partial}{\partial k^{\rho}}\,
\Pi_{\mu\nu\lambda\sigma}^{(\pi^0)}(q_1,q_2,k-q_1-q_2)
\nonumber\\
& & 
\!\!\!\!\!\!\!\!\!\!\!\!\!\!\!\!\!\!\!\!\!\!\!\!\!\!\!\!\!\!
=\,i\,
{\FF(q_1^2, q_2^2) \ \FF((q_1 + q_2)^2, 0) \over (q_1+q_2)^2 - 
M_{\pi}^2} \ \varepsilon_{\mu\nu\alpha\beta} \, q_1^\alpha q_2^\beta \ 
\varepsilon_{\lambda\sigma\rho\tau} \,  (q_1 + q_2)^\tau
\nonumber \\ 
& & 
\!\!\!\!\!\!\!\!\!\!\!\!\!\!\!\!\!\!\!\!\!\!\!\!\!\!\!
+\ i\, {\FF(q_1^2, 0) \ \FF(q_2^2, (q_1 + q_2)^2) \over
q_1^2 - M_{\pi}^2} \ \varepsilon_{\mu\sigma\tau\rho} \,
q_1^\tau \ \varepsilon_{\nu\lambda\alpha\beta} \,
q_1^\alpha q_2^\beta \nonumber \\
& & 
\!\!\!\!\!\!\!\!\!\!\!\!\!\!\!\!\!\!\!\!\!\!\!\!\!\!\!
+\ i\, {\FF(q_1^2, (q_1 + q_2)^2) \ \FF(q_2^2, 0) \over
q_2^2 - M_{\pi}^2} \ \varepsilon_{\mu\lambda\alpha\beta} \,
q_1^\alpha q_2^\beta \  \varepsilon_{\nu\sigma\rho\tau} \, 
q_2^\tau \nonumber \\
& &
\!\!\!\!\!\!\!\!\!\!\!\!\!\!\!\!\!\!\!\!\!\!\!\!\!\!\!
+\ \order(k)\,.
\lbl{derPi}   
\eea
Inserting this last expression into Eq.~\rf{Gamma2} and computing the
corresponding Dirac traces (we used {\small{REDUCE}}
\cite{hearn}) in Eq. \rf{F2trace}, we obtain
\bea
a_{\mu}^{\mbox{\tiny{LbyL;$\pi^0$}}}& = & - e^6 
\int {d^4 q_1 \over (2\pi)^4} \int {d^4 q_2 \over (2\pi)^4} 
\,\frac{1}{q_1^2 q_2^2 (q_1 + q_2)^2[(p+ q_1)^2 - m^2][(p - q_2)^2 - m^2]} 
\nonumber \\
&& \quad \quad \times \left[ 
{\FF(q_1^2, (q_1 + q_2)^2) \ \FF( q_2^2, 0) \over q_2^2 - 
M_{\pi}^2} \ T_1(q_1,q_2;p) \nonumber \right. \\ 
&& \quad \quad \quad + \left. {\FF( q_1^2,  q_2^2) \ \FF( (q_1 + q_2)^2,
0) \over (q_1 + q_2)^2 - M_{\pi}^2} \ T_2(q_1,q_2;p) \right] \,, 
\lbl{a_pion_2}  
\eea
with 
\bea
T_1(q_1,q_2;p) & = & {16 \over 3}\, (p \cdot q_1) \, (p \cdot q_2) \,
(q_1 \cdot q_2) 
\,-\, {16 \over 3}\, (p \cdot q_2)^2 \, q_1^2 \nonumber \\
&& \!\!\!\!\!
-\, {8 \over 3}\, (p \cdot q_1) \, (q_1 \cdot q_2) \, q_2^2 
\,+\, 8 (p \cdot q_2) \, q_1^2 \, q_2^2 
\,-\,{16 \over 3} (p \cdot q_2) \, (q_1 \cdot q_2)^2 \nonumber
\\
&&\!\!\!\!\!
+\, {16 \over 3}\, m^2 \, q_1^2 \, q_2^2 
\,-\, {16 \over 3}\, m^2 \, (q_1 \cdot q_2)^2 \, , \\
T_2(q_1,q_2;p) & = & {16 \over 3}\, (p \cdot q_1) \, (p \cdot q_2) \,
(q_1 \cdot q_2) \,-\,{16 \over 3}\, (p \cdot q_1)^2 \, q_2^2 
\nonumber \\
&&\!\!\!\!\!
 +\, {8 \over 3}\, (p \cdot q_1) \, (q_1 \cdot q_2) \, q_2^2 
\,+\, {8 \over 3}\, (p \cdot q_1) \, q_1^2 \, q_2^2 
\,\nonumber \\
&&\!\!\!\!\!
 +\, {8 \over 3}\, m^2 \, q_1^2 \, q_2^2   
\,-\, {8 \over 3}\, m^2 \, (q_1 \cdot q_2)^2 \, .     
\eea
In deriving Eq. \rf{a_pion_2}, we have used the fact that, upon a
trivial change of variables in the two-loop integral \rf{Gamma2}, the
two first terms of Eq.~\rf{derPi} lead to identical contributions.
Furthermore, in writing $T_2(q_1,q_2;p)$ we have taken into account
the invariance of the remaining factors of the corresponding integrand
under the exchange $q_1\leftrightarrow -q_2$.

Before discussing the pion-photon-photon transition form factor, let
us briefly mention a few features of Eq.~\rf{a_pion_2} that are
independent of the precise form of $\FF$. One observes that there are
five independent variables $p\cdot q_1, p \cdot q_2, q_1^2, q_2^2,$
and $q_1 \cdot q_2$, whereby $p \cdot q_i$ only occur in the fermion
propagators, not in the form factors. It might therefore be possible
to perform these two integrations over $p \cdot q_1$ and $p \cdot q_2$
for general form factors. We would then be left with a
three-dimensional integral, since the form factors depend on $q_1
\cdot q_2$.  Note that in both contributions in Eq.~\rf{a_pion_2}
there is one factor $\FF(q^2,0)$. In principle, this function could be
extracted, over a finite energy range, from the data collected, e.g., by
the CLEO collaboration~\cite{CLEO}. This would reduce the
model dependence.

\section{The pion-photon-photon transition form factor}
\label{sec:formfactors}
\renewcommand{\theequation}{\arabic{section}.\arabic{equation}}
\setcounter{equation}{0}

In the present work, we shall consider a representation of the form
factor $\FF$, based on the large-$N_C$ approximation to QCD, that
takes into account constraints from chiral symmetry at low energies,
and from the operator product expansion at short distances. Our
numerical estimates for $g-2$ below will be mainly based on
expressions for the form factor $\FF$ that involve either one vector
resonance (lowest meson dominance, LMD) or two vector resonances
(LMD+V). These {\it Ans\"atze} have been thoroughly discussed in
Ref.~\cite{paper_VAP}, and a short summary can be found in the
Appendix.  Furthermore, as a reference point we shall also consider
the simplest model for the form factor that follows from the
Wess-Zumino-Witten (WZW) term~\cite{WZW} that describes the
Adler-Bell-Jackiw anomaly~\cite{ABJ} in chiral perturbation
theory. Since in this case the form factor is constant, one needs an
ultraviolet cutoff, at least in the contribution to Eq.~\rf{a_pion_2}
involving $T_1$. Therefore, this model cannot be used for a reliable
estimate, but serves only illustrative purposes and as a check of our
calculation.  Finally we shall use the usual vector meson dominance
form factor. Essentially, this corresponds to the form factor that has
been employed in previous calculations for the pion-exchange
contribution to $g-2$ of the muon~\cite{KNO_85,HKS_95_96,HK_98}; see
also Ref.~\cite{Bijnens_Persson}. This will allow us to check the
numerics.  The expressions for the form factor $\FF$ based on the ENJL
model that have been used in Ref.~\cite{BPP} do not allow a
straightforward analytical calculation of the loop integrals. However,
compared with the results obtained in Refs.~\cite{HKS_95_96,HK_98},
the corresponding numerical estimates are rather close to the VMD case
(within the error attributed to the model dependence).  We would like
to stress again that both the VMD and ENJL form factors fail to
correctly reproduce the QCD short-distance constraints discussed
in~\cite{paper_VAP}. As noted earlier, the question of how sensitive
the results are to the latter motivated this work.

For the four cases mentioned above, the form factors are given
by~\cite{paper_VAP,Bijnens_Persson} [different choices for the global
sign in the normalization of the form factors are of no relevance,
cf. Eq. \rf{a_pion_2}]
\bea
\FF^{WZW}(q_1^2,q_2^2) & = & - {N_C \over 12 \pi^2 F_\pi} \, ,
\lbl{FF_WZW} \\ 
\FF^{VMD}(q_1^2,q_2^2) & = & - {N_C \over 12 \pi^2 F_\pi} {M_V^2 \over 
(q_1^2 - M_V^2)} {M_V^2 \over (q_2^2 - M_V^2)} \, , 
\lbl{FF_VMD}\\
\FF^{LMD}(q_1^2,q_2^2) & = &  {F_\pi \over 3}\, { q_1^2 + q_2^2 - c_V 
\over (q_1^2 - M_V^2) (q_2^2 - M_V^2) }  \, , 
\lbl{FF_LMD}\\
\FF^{LMD+V}(q_1^2,q_2^2) & = &  {F_\pi \over 3}\, { q_1^2 q_2^2 
(q_1^2 + q_2^2) + h_1 (q_1^2 + q_2^2)^2 + h_2 q_1^2 q_2^2 + h_5
(q_1^2 + q_2^2) + h_7 \over (q_1^2 - M_{V_1}^2) (q_1^2 - M_{V_2}^2)
(q_2^2 - M_{V_1}^2) (q_2^2 - M_{V_2}^2)} \, , \lbl{FF_LMD+V} 
\eea
with (see the Appendix)
\be
c_V \,= \, {N_C \over 4\pi^2} {M_V^4 \over F_\pi^2} \, , 
\quad h_7 \,= \, - {N_C \over 4\pi^2} {M_{V_1}^4 M_{V_2}^4 \over
F_\pi^2} \, . 
\ee 

According to Ref.~\cite{Brodsky_Lepage}, the form factor $\FF(-Q^2,0)$
with one photon on shell behaves like $1/Q^2$ for large spacelike
momenta, $Q^2=-q^2$. The experimental data are compatible with this
behavior~\cite{CLEO}. Whereas the LMD form factor does not have such a
behavior, it can be reproduced with the LMD+V {\it Ansatz}, provided
that $h_1 = 0$. In Ref.~\cite{paper_VAP}, a fit of the LMD+V form
factor to the CLEO data yielded $h_1 = -0.01 \pm 0.16~\mbox{GeV}^2,\,
h_5 = 6.88 \pm 0.61~\mbox{GeV}^4$, or, if we fit $h_5$ with $h_1 = 0$,
$h_5 = 6.93 \pm 0.26~\mbox{GeV}^4$. On the other hand, the parameter
$h_2$ that appears in Eq.~\rf{FF_LMD+V} is not known. If we compare,
in the context of Ref.~\cite{Pi_ll}, the expression for the $\pi^0 \to
e^+ e^-$ partial width based on the LMD+V form factor to the
experimental values \cite{alavi99,PDG00}, we obtain a rather loose
constraint on $h_2$, $|h_2| \lapprox 20~\mbox{GeV}^2$, although
slightly positive values for $h_2$ seem to be preferred in order to
reproduce the experimental rate.

A crucial observation at this stage is that {\it all} dependences on
$q_1 \cdot q_2$ in the numerators in $\FF^{LMD+V,LMD}(q_1^2, (q_1 +
q_2)^2)$ can be canceled by the ones in the resonance
propagators. Therefore, we can write
\be \lbl{FF_f_g}
\FF(q_1^2, q_2^2) \,=\, \frac{F_\pi}{3}\,\bigg[ f(q_1^2)\,-\, 
\sum_{M_{V_i}} {1 \over q_2^2 - M_{V_i}^2}
g_{M_{V_i}}(q_1^2)\bigg] \, . 
\ee
This representation, which seems to hold quite generally in the
large-$N_C$ limit of QCD (see the Appendix), obviously extends to the
VMD and WZW form factors. The corresponding functions $f(q^2)$ and
$g_{M_{V_i}}(q^2)$ can easily be worked out from the explicit
expressions \rf{FF_WZW}--\rf{FF_LMD+V}.  They are displayed in
Table~\ref{tab:tab1}.  For the VMD and LMD form factors, the sum in
Eq.~\rf{FF_f_g} reduces to a single term, and we call the corresponding
function $g_{M_V}(q^2)$.  In the case of the LMD+V form factor, we
have written
\bea
g_{M_{V_1}}(q^2) &=& \frac{1}{M_{V_2}^2 - M_{V_1}^2}\,
\frac{g(q^2;M_{V_1}^2)}{(q^2-M_{V_1}^2)(q^2-M_{V_2}^2)}\,,
\nonumber\\
\nonumber\\
g_{M_{V_2}}(q^2) &=& \frac{1}{M_{V_1}^2 - M_{V_2}^2}\,
\frac{g(q^2;M_{V_2}^2)}{(q^2-M_{V_1}^2)(q^2-M_{V_2}^2)}\,,
\lbl{gLMDV}
\eea
with the function $g(q^2;M^2)$ given in the table.  The
expression~\rf{FF_f_g} also holds for $q_1^2 = 0$, which implies that
$f(0) = 0$ in order to reproduce the asymptotic $1/Q^2$ behavior of
$\FF(-Q^2,0)$. As noted above, this fails to be the case for the LMD
{\it Ansatz} $f^{LMD}(0) \neq 0$, and holds for the LMD+V {\it Ansatz}
provided $h_1=0$.  Note that the limit $Q_2^2 \to \infty$ with $Q_1^2$
fixed can actually not be studied with the operator product expansion
(see the discussion in Ref.~\cite{paper_VAP}). 

\begin{table}
\caption{The functions $f(q^2)$, $g_{M_V}(q^2)$, and $g(q^2;M^2)$ of Eqs.
\protect\rf{FF_f_g} and \protect\rf{gLMDV} for the different form factors in 
Eqs. \protect\rf{FF_WZW}--\protect\rf{FF_LMD+V}.}
\begin{center}
\renewcommand{\arraystretch}{1.1}
\begin{tabular}{|c|c|c|}
\hline
 & $f(q^2)$ & $g_{M_V}(q^2)$ \\[0.1cm]
\hline
\rule[0mm]{0mm}{1cm} 
$WZW$ & $\displaystyle-\,{\frac{N_C}{4\pi^2F_{\pi}^2}}$  & 0
\\[0.4cm]
$VMD$ & 0 & $\displaystyle{\frac{N_C}{4\pi^2F_{\pi}^2}\,
\frac{M_V^4}{q^2-M_V^2}}$ 
\\[0.4cm]
$LMD$ & $\displaystyle{\frac{1}{q^2-M_V^2}}$ & $\displaystyle{-\,
\frac{q^2+M_V^2-c_V}{q^2-M_V^2}}$
\\[0.4cm]
$LMD+V$ & $\displaystyle{\frac{q^2+h_1}{(q^2-M_{V_1}^2)(q^2-M_{V_2}^2)}}$ & 
$g(q^2;M^2) =  q^2 M^2 (q^2+M^2)+h_1(q^2+M^2)^2$
\\
\rule[-0.5cm]{0mm}{0.5cm} 
&   &  $ \qquad\qquad \ \  +\, h_2q^2M^2 + h_5(q^2+M^2) + h_7$
\\
\hline
\end{tabular}
\label{tab:tab1} 
\end{center}
\end{table}

Thus, using partial fractions, we can write the product of the form
factors in the two contributions to the right-hand side of
Eq.~\rf{a_pion_2} as follows:
\bea
{\FF(q_1^2, (q_1 + q_2)^2) \ \FF(q_2^2, 0) \over q_2^2 - 
M_{\pi}^2} & = & \frac{F_\pi}{3}\,{\FF(q_2^2, 0) \over q_2^2 -
M_{\pi}^2} \Bigg( 
f(q_1^2) \nonumber \\
&& \quad - \sum_{M_{V_i}} {g_{M_{V_i}}(q_1^2)\over (q_1 + q_2)^2 - M_{V_i}^2} 
 \Bigg)  \lbl{prefactor_T1} 
\eea
and 
\bea 
\lefteqn{ {\FF( q_1^2, q_2^2) \ \FF( (q_1 + q_2)^2, 0) \over (q_1
+ q_2)^2 - M_{\pi}^2} = \frac{F_\pi}{3}\,\FF(q_1^2, q_2^2) } \nonumber \\
&& \qquad \qquad\qquad\qquad \times \left( {1 \over (q_1 + q_2)^2 - 
M_\pi^2} \left[ f(0) + \sum_{M_{V_i}} { g_{M_{V_i}}(0) \over
(M_{V_i}^2 - M_\pi^2) }  \right] \right. \nonumber \\
&& \qquad\qquad\qquad\qquad\qquad  \left. + \sum_{M_{V_i}} 
{g_{M_{V_i}}(0)\over [(q_1 + q_2)^2 - M_{V_i}^2]  (M_\pi^2 - M_{V_i}^2)}
\right) \, . \lbl{prefactor_T2} 
\eea
Note that in Eq.~\rf{prefactor_T2} all terms contain a propagator $1 /
[(q_1+q_2)^2 - M^2]$, $M=M_\pi$ or $M_V$, in contrast to the prefactor
of $T_1$ shown in Eq.~\rf{prefactor_T1}.

Since the form factors discussed above go to a constant for small
momenta, the integrals in Eq. \rf{a_pion_2} are infrared finite.  The
behavior at large momenta also ensures the ultraviolet convergence of
these integrals, except, of course, for the constant form factor
$\FF^{WZW}$.  However, the integral involving $T_2(q_1,q_2;p)$ is
finite in the latter case also (this has been observed before
\cite{HK_98}, and we have confirmed it numerically as well).

\section{Angular integrations}
\label{sec:angular} 
\renewcommand{\theequation}{\arabic{section}.\arabic{equation}}
\setcounter{equation}{0}

As we shall show in this section, for form factors $\FF(q_1^2,q_2^2)$
that have the general form given by Eq.~\rf{FF_f_g}, it is possible to
perform {\it all} angular integrations in the two-loop integral of
Eq.~\rf{a_pion_2} using the technique of Gegenbauer polynomials
(hyperspherical approach); see Refs.~\cite{early_Gegenbauer,LRR,RRL}.
In order to do so, we perform a Wick rotation of the momenta
$q_1^{\mu}$, $q_2^{\mu}$, and $p^{\mu}$, denoting by capital letters
the rotated Euclidean momenta, with $Q_i^2=-q_i^2$, $P^2=-m^2$,
etc. Let us mention that in the language of Refs.~\cite{LRR,RRL} the
loop integrals we have to deal with are planar.  We also note that
these integrals are of the two-loop self-energy type. Other techniques
that lead to either one-dimensional dispersive or two-dimensional
integral representations~\cite{master} could {\it a priori} also be
relevant in the present context, although probably not without further
specifying the functions $f(q^2)$ and $g_{M_{V_i}}(q^2)$ in
Eq.~\rf{FF_f_g}.

\subsection{Properties of Gegenbauer polynomials}

Let us briefly summarize some basic properties of the Gegenbauer
polynomials; see also Refs.~\cite{MOS,AS}.  We write the measure of
the four-dimensional sphere as follows (Euclidean momenta): 
\be
d^4 K = K^3 dK \, d\Omega(\hat K)\, , \quad 
\int d\Omega(\hat K) = 2 \pi^2 \, . 
\ee
The generating function of the Gegenbauer polynomials\footnote{We
shall need only the special case $C_n(x) \equiv 
C_n^{(1)}(x)$.} $C_n(x)$ is given by
\be \lbl{gen_func} 
{1 \over z^2 - 2 x z + 1} = \sum_{n=0}^\infty z^n \, C_n(x) \, , \quad -1
\leq x \leq 1, \ |z| < 1\, .
\ee
From Eq.~\rf{gen_func} we immediately obtain the following property
under parity transformations $C_n(-x) = (-1)^n C_n(x)$. Furthermore we 
get $C_n(1) = n + 1$. The Gegenbauer polynomials obey the
orthogonality conditions  
\bea
\int d\Omega(\hat K) \, C_n(\hat Q_1 \cdot \hat K) \, C_m(\hat K \cdot
\hat Q_2) & = & 2 \pi^2 {\delta_{nm} \over n + 1} C_n(\hat Q_1 \cdot \hat 
Q_2) \, , \nonumber \\ 
\int d\Omega(\hat K) \, C_n(\hat Q \cdot \hat K) \, C_m(\hat K \cdot
\hat Q)  & = & 2 \pi^2 \delta_{nm} \, , 
\eea
where, for instance, $\hat Q_1 \cdot \hat K$ is the cosine of the angle
between the four-dimensional vectors $Q_1$ and $K$.  Some low-order
cases of the polynomials are given by $C_0(x) = 1, \ C_1(x) = 2x,
\ C_2(x) = 4 x^2 - 1$ and therefore $x = C_1(x)/2, \ x^2 = [
C_2(x) + C_0(x) ] / 4$.

{F}rom the generating function, we obtain the following representation
of the propagators in Euclidean space: 
\bea
{1 \over (K-L)^2 + M^2} & = & {Z_{KL}^M \over |K| |L|}
\sum_{n=0}^\infty (Z_{KL}^M)^n \, C_n(\hat K \cdot \hat L) \, , \\ 
{1 \over (K+L)^2 + M^2} & = & {Z_{KL}^M \over |K| |L|}
\sum_{n=0}^\infty (- Z_{KL}^M)^n \, C_n(\hat K \cdot \hat L) \, , \\ 
Z_{KL}^M & = & {K^2 + L^2 + M^2 - \sqrt{(K^2 + L^2 + M^2)^2 -
4 K^2 L^2} \over 2 |K| |L|} \, . 
\eea
Note that we have to choose the negative sign in front of the square
root in $Z_{KL}^M$ in order that $|Z_{KL}^M| < 1$. For a massless
propagator these expressions simplify as follows: 
\bea
{1 \over (K-L)^2} & = & \theta\left(1 - {|L| \over |K|}\right) \, {1 \over
K^2} \, \sum_{n=0}^\infty \left( {|L| \over |K|} \right)^n C_n(\hat K
\cdot \hat L) \nonumber \\ 
&&+ \, \theta\left(1 - {|K| \over |L|}\right) \, {1 \over L^2} \, 
\sum_{n=0}^\infty \left( {|K| \over |L|} \right)^n C_n(\hat K \cdot
\hat L) \, . 
\eea

\subsection{Basic angular integrals}

Let us introduce the following abbreviations for the propagators in
the loop integral in Eq.~\rf{a_pion_2}:
\bea
D_1 & = & Q_1^2 \, , \quad 
D_2 = Q_2^2 \, , \quad 
D_3 = (Q_1 + Q_2)^2 \, , \quad 
D_4 = (P+Q_1)^2 + m^2 \, , \nonumber \\ 
D_5 & = & (P-Q_2)^2 + m^2 \, , \quad
D_M = (Q_1 + Q_2)^2 + M^2 \, , 
\eea
where $M$ denotes either the pion mass or the mass of some vector
resonance. For $M \to 0$ we recover the photon propagator $D_3$.

By rewriting the scalar products $P \cdot Q_1, P \cdot Q_2,$ and $Q_1
\cdot Q_2$ in terms of the propagators $D_4, D_5,$ and $D_3$,
respectively, we can remove  some of the terms in the numerator of
$T_n/(D_1 D_2 D_3 D_4 D_5)$, $n=1,2$, in Eq.~\rf{a_pion_2}. If we
multiply by the form factors, taking into account their general form
from Eqs.~\rf{prefactor_T1} and \rf{prefactor_T2}, and use a
partial fraction decomposition, we finally obtain the following basic
angular integrals [apart from the trivial one $\int d\Omega(\hat Q_1)
d\Omega(\hat Q_2) = 4 \pi^4$], which can all be performed using the
method of Gegenbauer polynomials (from now on, we write $Q_1$ instead of 
$|Q_1|$, etc.):
\bea
I_1^M & = & \int {d\Omega(\hat Q_1) \over 2\pi^2} {d\Omega(\hat Q_2)
\over 2\pi^2} \, {1 \over D_M D_4 D_5} =  - {1 \over P^2 Q_1^2 Q_2^2} \, 
\ln\left( 1 - Z_{Q_1 Q_2}^M \, Z_{P Q_1}^m \, Z_{P Q_2}^m \right) \, ,
\nonumber \\  
& = & {1 \over m^2 Q_1^2 Q_2^2} \ln \left[ 1 + {
(M^2 + Q_1^2 + Q_2^2 - R^M) \, (Q_1^2 - R_1^m) \, (Q_2^2 - R_2^m) \over 8
m^2 Q_1^2 Q_2^2 } \right] \, , \nonumber \\
I_2 & = & \int {d\Omega(\hat Q_1) \over 2\pi^2} {d\Omega(\hat Q_2)
\over 2\pi^2} {1 \over D_4 D_5} = {Z_{P Q_1}^m \, Z_{P
Q_2}^m \over P^2  \, Q_1 \, Q_2} \, , \nonumber \\
& = & {(Q_1^2 - R_1^m) \, (Q_2^2 - R_2^m) \over 4 m^4 Q_1^2
Q_2^2} \, , \nonumber \\
I_3^M & = & \int {d\Omega(\hat Q_1) \over 2\pi^2} {d\Omega(\hat Q_2)
\over 2\pi^2} {1 \over D_M D_4} = {Z_{Q_1 Q_2}^M \, Z_{P
Q_1}^m \over P \, Q_2 \, Q_1^2}  \, , \nonumber \\
& = & - { (M^2 + Q_1^2 + Q_2^2 - R^M) \, (Q_1^2 - R_1^m) \over
4 m^2 Q_1^4 Q_2^2} \, , \nonumber \\
I_4^M & = & \int {d\Omega(\hat Q_1) \over 2\pi^2} {d\Omega(\hat Q_2)
\over 2\pi^2} {1 \over D_M D_5} = {Z_{Q_1 Q_2}^M \, Z_{P Q_2}^m \over
P  \, Q_1 \, Q_2^2} \, , \nonumber \\
& = & - { (M^2 + Q_1^2 + Q_2^2 - R^M) \, (Q_2^2 - R_2^m) \over
4 m^2 Q_1^2 Q_2^4} \, , \nonumber \\
I_5 & = & \int {d\Omega(\hat Q_1) \over 2\pi^2} {d\Omega(\hat Q_2)
\over 2\pi^2} {1 \over D_4} = {Z_{P Q_1}^m \over P \,
Q_1} \, , \nonumber \\ 
& = & - {Q_1^2 - R_1^m \over 2 m^2 Q_1^2} \, , \nonumber \\
I_6 & = & \int {d\Omega(\hat Q_1) \over 2\pi^2} {d\Omega(\hat Q_2)
\over 2\pi^2} {1 \over D_5} = {Z_{P Q_2}^m \over P \, Q_2} \, ,
\nonumber  \\
& = & - {Q_2^2 - R_2^m \over 2 m^2 Q_2^2} \, , \nonumber \\
I_7^M & = & \int {d\Omega(\hat Q_1) \over 2\pi^2} {d\Omega(\hat Q_2)
\over 2\pi^2} {1 \over D_M} = {Z_{Q_1 Q_2}^M \over Q_1 \, Q_2} \,
, \nonumber \\ 
& = & {M^2 + Q_1^2 + Q_2^2 - R^M \over 2 Q_1^2 Q_2^2} \, , 
\eea
where
\bea
R^M & = & \sqrt{(M^2 + Q_1^2 + Q_2^2)^2 - 4 Q_1^2 Q_2^2} \, ,
\nonumber \\
R_i^m & = & \sqrt{Q_i^4 + 4 m^2 Q_i^2} \, , \quad  i = 1,2 \, .  
\eea

We shall also need the angular integrals where the massive propagator
$D_M$ is replaced by the massless one $D_3$.  We have set $P^2 = -
m^2$ in the explicit expressions involving the square roots.  Since
the external momentum $P$ flows only through the massive fermion
propagators, we do not need to deform the integration contour for the
radial integrals over $Q_1^2$ and $Q_2^2$ (see the discussion in
Refs.~\cite{LRR,RRL}). Note that $I_5$ and $I_6$ depend only on one
variable, $Q_1^2$ and $Q_2^2$, respectively, whereas the expression
for $I_2$ factorizes into a product of two single-variable functions.

Other angular integrals that occur during the calculation, involving
the factors $D_3 /(D_4 D_5)$, $D_4 / (D_M D_5)$, $D_5 / (D_M D_4),$
and $D_3 / D_4$, can be reduced to the basic ones, by noting that $D_3
= Q_1^2 + Q_2^2 + 2 Q_1 \cdot Q_2$, $D_4 = Q_1^2 + 2 P \cdot Q_1,$ and
$D_5 = Q_2^2 - 2 P \cdot Q_2$, and with the help of the identities
\bea
\left( Z_{Q_1 Q_2}^M \right)^2 & = & {M^2 + Q_1^2 + Q_2^2 \over Q_1
Q_2} \, Z_{Q_1 Q_2}^M - 1 \, , \nonumber \\
\left( Z_{P Q_i}^m \right)^2 & = & {m^2 + P^2 + Q_i^2 \over P Q_i}
\, Z_{P Q_i}^m - 1 \, . 
\eea
In this way one obtains, for $P^2 = - m^2$,
\bea
\int {d\Omega(\hat Q_1) \over 2\pi^2} {d\Omega(\hat Q_2) \over 2\pi^2}
\, {Q_1 \cdot Q_2 \over D_4 D_5} & = & - {1 \over 4} {\left( Z_{P 
Q_1}^m  \right)^2 \left( Z_{P Q_2}^m \right)^2 \over P^2} \, ,
\nonumber \\ 
& = & { Q_1^2 Q_2^2 \, I_2 - Q_1^2 \, I_5 - Q_2^2 \, I_6 + 1
\over 4 m^2} \, , \nonumber \\ 
\int {d\Omega(\hat Q_1) \over 2\pi^2} {d\Omega(\hat Q_2) \over 2\pi^2}
\, {P \cdot Q_1 \over D_M D_5} & = & - {1 \over 4} {\left( Z_{Q_1 
Q_2}^M \right)^2 \left( Z_{P Q_2}^m \right)^2 \over Q_2^2} \, ,
\nonumber \\
& = & { - [M^2 + Q_1^2 + Q_2^2] Q_2^2 \, I_4^M + (M^2 + Q_1^2 + Q_2^2)
\, I_7^M + Q_2^2 \, I_6 - 1 \over 4 Q_2^2} \, , \nonumber \\     
\int {d\Omega(\hat Q_1) \over 2\pi^2} {d\Omega(\hat Q_2) \over 2\pi^2}
\, {P \cdot Q_2 \over D_M D_4} & = & {1 \over 4} {\left( Z_{Q_1 Q_2}^M
\right)^2 \left( Z_{P Q_1}^m \right)^2 \over Q_1^2} \, , \nonumber \\ 
& = & { [M^2 + Q_1^2 + Q_2^2] Q_1^2
\, I_3^M - (M^2 + Q_1^2 + Q_2^2) \, I_7^M - Q_1^2 \, I_5 + 1 \over 4
Q_1^2}  \, , \nonumber \\
\int {d\Omega(\hat Q_1) \over 2\pi^2} {d\Omega(\hat Q_2) \over 2\pi^2}
\, {Q_1 \cdot Q_2 \over D_4} & = & 0 \, .  
\eea

\section{Two-dimensional integral representation for 
$a_{\mu}^{\mbox{\small{LbyL;$\pi^0$}}}$} 
\label{sec:twodimrepresentation}
\renewcommand{\theequation}{\arabic{section}.\arabic{equation}}
\setcounter{equation}{0}

After having performed the angular integrations, the pion-exchange
contribution to $g-2$ can be written in a two-dimensional integral
representation as follows:
\bea
a_{\mu}^{\mbox{\tiny{LbyL;$\pi^0$}}} & = & 
\left( {\alpha \over \pi } \right)^3 
\left[ a_{\mu}^{\mbox{\tiny{LbyL;$\pi^0$}}(1)} + 
a_{\mu}^{\mbox{\tiny{LbyL;$\pi^0$}}(2)} \right] \, , \lbl{api_two_dim} \\
a_{\mu}^{\mbox{\tiny{LbyL;$\pi^0$}}(1)} 
& = & \int_0^\infty dQ_1 \int_0^\infty dQ_2 \ \Bigg[
w_{f_1}(Q_1,Q_2) \ f^{(1)}(Q_1^2,Q_2^2) \nonumber \\ && \qquad \qquad
\qquad \qquad + \sum_{M_{V_i}} \ w_{g_1}(M_{V_i},Q_1,Q_2) \
g_{M_{V_i}}^{(1)}(Q_1^2, Q_2^2) \Bigg] \, , \lbl{api1} \\
a_{\mu}^{\mbox{\tiny{LbyL;$\pi^0$}}(2)} 
& = & \int_0^\infty dQ_1 \int_0^\infty dQ_2 \ 
\sum_{M=M_\pi, M_{V_i}} \ w_{g_2}(M,Q_1,Q_2) \ g_{M}^{(2)}(Q_1^2,
Q_2^2) 
\, , \lbl{api2}
\eea
with [see Eqs.~\rf{prefactor_T1}--\rf{prefactor_T2}] 
\bea
f^{(1)}(Q_1^2,Q_2^2) & = & \frac{F_\pi}{3\,}f(-Q_1^2) \ \FF(-Q_2^2,0) 
 \, , \nonumber \\ 
g_{M_{V_i}}^{(1)}(Q_1^2,Q_2^2) & = & \frac{F_\pi}{3}\,
{g_{M_{V_i}}(-Q_1^2) \over M_{V_i}^2} \ \FF(-Q_2^2,0) \, , \nonumber \\ 
g_{M_\pi}^{(2)}(Q_1^2,Q_2^2) & = & \frac{F_\pi}{3}\,\FF(-Q_1^2,-Q_2^2)
\ \left( f(0) + \sum_{M_{V_i}} {g_{M_{V_i}}(0) \over M_{V_i}^2 - M_\pi^2 }
\right) \, , \nonumber \\
g_{M_{V_i}}^{(2)}(Q_1^2,Q_2^2) & = & \frac{F_\pi}{3}\,\FF(-Q_1^2,-Q_2^2) \ 
{g_{M_{V_i}}(0) \over M_\pi^2 - M_{V_i}^2} \, . 
\lbl{hatf1_hatg2M}
\eea
Note (see Table~\ref{tab:tab1}) that in the WZW model we obtain
$g_{M_{V_i}}^{(1)}(Q_1^2,Q_2^2) \equiv 0$,
$g_{M_{V_i}}^{(2)}(Q_1^2,Q_2^2) \equiv 0$, since $g^{WZW}(-Q_1^2) 
\equiv 0$. On the other hand, for the VMD
model we have the simplifications $f^{(1)}(Q_1^2,Q_2^2) \equiv 0$,
$g_{M_{V_1}}^{(2)}(Q_1^2,Q_2^2) = - g_{M_\pi}^{(2)}(Q_1^2,Q_2^2)$,
since $f^{VMD}(-Q_1^2) \equiv 0$.

The universal [for the class of form factors that have a
representation of the type \rf{FF_f_g}] weight functions in
Eqs.~\rf{api1} and \rf{api2} are given by 
\bea
w_{f_1}(Q_1,Q_2)
& = & {\pi^2 \over Q_2^2 + M_\pi^2}\,\frac{1}{6 m^2 Q_1 Q_2}\, \ \Bigg\{
- 4 (2 m^2 - Q_2^2) (Q_1^2 - Q_2^2)^2 
\nonumber \\ 
&&\qquad \times \ln \left[ 1 + { (Q_1^2 +
Q_2^2 - R^0) (Q_1^2 - R_1^m) (Q_2^2 - R_2^m) \over 8 m^2 Q_1^2 Q_2^2 }
\right]
\nonumber \\ 
&& + \bigg[ - 4m^2 Q_1^2 Q_2^2 -
{1 \over m^2} Q_1^4 Q_2^4 + {1 \over 2 m^4} Q_1^4 Q_2^6 + 
Q_1^6 
\nonumber \\
&&
\qquad 
- 3 Q_1^4 Q_2^2 - Q_1^2 Q_2^4 + Q_2^6
\bigg]\bigg(1\,-\,\frac{R_1^m}{Q_1^2}\bigg)  
\nonumber\\ 
&& 
-(Q_1^2-Q_2^2)^2 R^0 \bigg(1\,-\,\frac{R_1^m}{Q_1^2}\bigg)
\nonumber\\
&& 
+ \,\frac{1}{m^2}\,Q_2^2 (Q_2^4-4m^4) R_1^m
-\,\frac{1}{2m^4}\,Q_1^4(Q_2^2-2m^2)^2 R_2^m 
\nonumber\\
&& 
+ \,\frac{1}{2m^4}\,(Q_2^2-2m^2)(Q_1^2Q_2^2 - 2m^2Q_1^2 - 2m^2Q_2^2)
R_1^m R_2^m  
 \Bigg\}
\, , 
\lbl{wf1} 
\\
w_{g_1}(M,Q_1,Q_2)
& =& {\pi^2 \over Q_2^2 + M_\pi^2}\,\frac{1}{6 m^2 Q_1 Q_2}\, \ \Bigg\{
 - 4 (2 m^2 - Q_2^2) (Q_1^2 - Q_2^2)^2 
\nonumber \\
&&\qquad \times \ln \left[ 1 + { (Q_1^2 + Q_2^2 - R^0) (Q_1^2 -
     R_1^m) (Q_2^2 - R_2^m) \over 8 m^2 Q_1^2 Q_2^2 } \right]
\nonumber \\ 
&&\, + 4 (2 m^2 - Q_2^2) [M^4 + (Q_1^2 - Q_2^2)^2 + 2 M^2
     (Q_1^2 + Q_2^2)] 
 \nonumber \\
&& \qquad \times 
     \ln \left[ 1 + { (M^2 + Q_1^2 + Q_2^2 - R^M) (Q_1^2 -
     R_1^m) (Q_2^2 - R_2^m) \over 8 m^2 Q_1^2 Q_2^2 } \right]
\nonumber \\ 
&&
-\, \bigg[
 M^6 + 3M^4Q_1^2 + 3M^2Q_1^4 + 3M^4Q_2^2 + 2M^2Q_1^2Q_2^2 + 3M^2Q_2^4 
\nonumber\\
&&
\qquad
-\,\frac{M^2}{m^2}\,Q_1^2Q_2^4 \bigg ]\bigg(1\,-\,\frac{R_1^m}{Q_1^2}\bigg)
\nonumber\\
&&
- (Q_1^2 - Q_2^2)^2 R^0 \bigg(1\,-\,\frac{R_1^m}{Q_1^2}\bigg)
\, - \,\frac{M^2}{m^2}\,Q_1^2(Q_2^2 - 2m^2)R_2^m
\bigg(1\,-\,\frac{R_1^m}{Q_1^2}\bigg) 
\nonumber\\
&&
+\,\bigg[ M^4 + 2M^2Q_1^2 + 2 M^2Q_2^2 + Q_1^4 + Q_2^4 - 2Q_1^2Q_2^2\bigg]
\nonumber\\
&&
\qquad\times
R^M\bigg(1\,-\,\frac{R_1^m}{Q_1^2}\bigg)
\Bigg\}
\, , \\
w_{g_2}(M,Q_1,Q_2)
& =&  \frac{\pi^2}{6 m^2 M^2 Q_1 Q_2}\, \ \Bigg\{
4 [m^2 (Q_2^2 - Q_1^2) + 2 Q_1^2 Q_2^2] (Q_1^2 - Q_2^2) 
\nonumber \\
&&\qquad \times 
     \ln \left[ 1 + { (Q_1^2 + Q_2^2 - R^0) (Q_1^2 -
     R_1^m) (Q_2^2 - R_2^m) \over 8 m^2 Q_1^2 Q_2^2 } \right]
\nonumber \\ 
&& + 4 \Big[ M^4 m^2 + (Q_1^2 - Q_2^2) [-2
    Q_1^2 Q_2^2 + m^2 (Q_1^2 - Q_2^2)] 
\nonumber \\
&& \qquad \qquad \qquad \quad + 2 M^2 [Q_1^2 Q_2^2 + m^2 (Q_1^2 + Q_2^2)]
      \Big] \nonumber \\
&&\qquad \times 
     \ln \left[ 1 + { (M^2 + Q_1^2 + Q_2^2 - R^M) (Q_1^2 -
     R_1^m) (Q_2^2 - R_2^m) \over 8 m^2 Q_1^2 Q_2^2 } \right]
\nonumber\\
&&
\,+ M^4Q_1^2 + 2M^2Q_1^4 + M^4Q_2^2 - M^2Q_1^2Q_2^2 + 2M^2Q_2^4
\nonumber \\ 
&&
+\, Q_1^2(Q_1^2 + Q_2^2) R^0 \bigg(1\,-\,\frac{R_1^m}{Q_1^2}\bigg)
\,+\,Q_2^2(Q_2^2 - 3Q_1^2) R^0 \bigg(1\,-\,\frac{R_2^m}{Q_2^2}\bigg)
\nonumber\\
&&
-\, Q_1^2(M^2 + Q_1^2 + Q_2^2) R^M \bigg(1\,-\,\frac{R_1^m}{Q_1^2}\bigg)
\,-\,M^2(M^2 + 2Q_1^2 + Q_2^2) R_1^m
\nonumber\\
&&
-\, Q_2^2(M^2 - 3Q_1^2 + Q_2^2) R^M \bigg(1\,-\,\frac{R_2^m}{Q_2^2}\bigg)
 \,-\, M^2(M^2 - 3Q_1^2 + 2Q_2^2) R_2^m
\nonumber\\
&& 
-\,M^2 R_1^m R_2^m \Bigg\}
\, , \lbl{wg2M} 
\eea
where
\be
R^0 \equiv R^{M=0} = \sqrt{(Q_1^2 + Q_2^2)^2 - 4 Q_1^2 Q_2^2} \, . 
\ee

\begin{figure}[!t]

\indent

\centerline{\psfig{figure=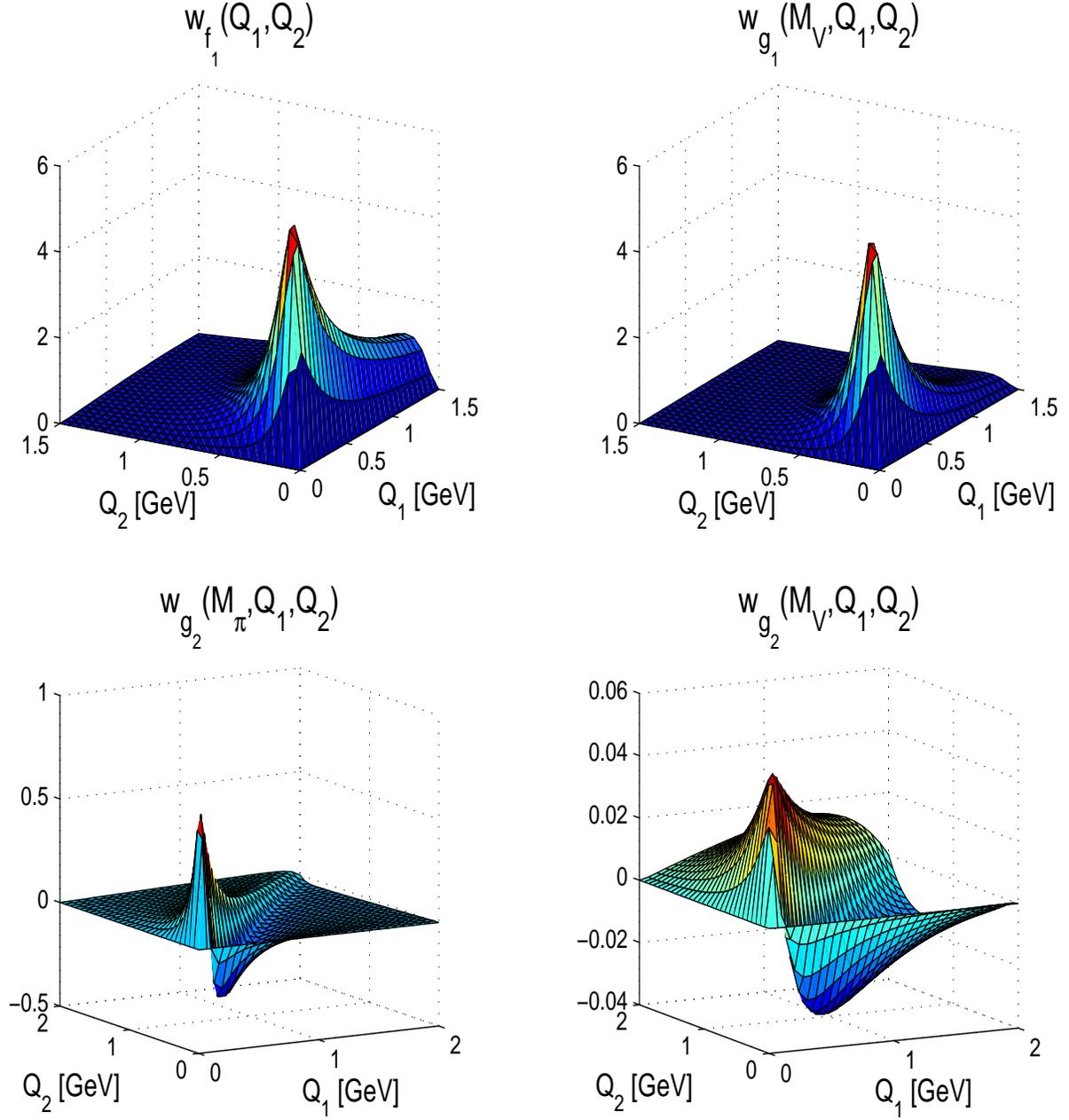,height=17cm,width=16cm}}

\indent

\caption{The weight functions of
Eqs. \protect\rf{wf1}--\protect\rf{wg2M}. Note the different 
ranges of $Q_i$ in the subplots.  The functions $w_{f_1}$ and
$w_{g_1}$ are positive definite and peaked in the region $Q_1\sim
Q_2\sim 0.5$ GeV. Note, however, the tail in $w_{f_1}$ in the
$Q_1$ direction for $Q_2 \sim 0.2~\mbox{GeV}$.  The functions
$w_{g_2}(M_\pi,Q_1,Q_2)$ and $w_{g_2}(M_V,Q_1,Q_2)$ take both signs,
but their magnitudes remain small as compared to $w_{f_1}(Q_1,Q_2)$
and $w_{g_1}(M_V,Q_1,Q_2)$.  We have used $M_V=M_{\rho}=769$ MeV.}
\label{fig:fig3}

\indent

\end{figure}

Although the analytical expressions for the weight functions look
quite complicated and involve terms with different signs and of
different sizes, the sum of all terms leads to rather smooth functions
of the two variables $Q_1$ and $Q_2$, as can be seen from the plots in
Fig.~\ref{fig:fig3}.  We have not shown the corresponding plots for
$w_{g_1}(M_{V_2},Q_1,Q_2)$ and $w_{g_2}(M_{V_2},Q_1,Q_2)$, since they
look qualitatively similar.  The contribution of the former is
suppressed in the integral by the factor $1/M_{V_2}^2$ in
$g_{M_{V_2}}^{(1)}$ whereas the latter weight function scales as 
$1/M_{V_2}^2$ for large $M_{V_2}$.

The functions $w_{f_1}$ and $w_{g_1}$ are positive and concentrated
around momenta of the order of $0.5~\mbox{GeV}$. This behavior was
already observed in Ref.~\cite{BPP} by varying the upper bound of the
integrals (an analogous analysis is contained in
Ref.~\cite{HKS_95_96}).  Note, however, the tail in $w_{f_1}$ in the
$Q_1$ direction for $Q_2 \sim 0.2~\mbox{GeV}$. On the other hand, the
function $w_{g_2}$ has positive and negative contributions in that
region, which will lead to a strong cancellation in the corresponding
integrals, provided they are multiplied by a positive function
composed of the form factors (see the numerical results below).

As can be seen from the plots, and checked analytically, the weight
functions vanish for small momenta. Therefore, as already noted
before, the integrals are infrared finite.  For large momenta the
weight function $w_{f_1}(Q_1,Q_2)$ and $w_{g_1}(M,Q_1,Q_2)$ have the
following behavior:
\bea
\lim_{Q_1\to\infty} w_{f_1}(Q_1,Q_2) & = &  
\pi^2 \, {[ Q_2^5 - Q_2 (Q_2^2 - 2 m^2) R_2^m] \over m^2 (Q_2^2 +
M_\pi^2) Q_1} + \order\left( {1 \over Q_1^3} \right) \, , 
\quad{\mbox{ $Q_2$ fixed}} \, ,  
\nonumber\\
\lim_{Q_2\to\infty} w_{f_1}(Q_1,Q_2) & = &  
\pi^2 \, { [ 2 Q_1^3 (Q_1^2 - 12 m^2) - 2 Q_1 (Q_1^2 - 14 m^2) R_1^m] 
\over 9 m^2 Q_2^3} + \order\left( {1 \over Q_2^5 } \right) \, , 
\quad{\mbox{ $Q_1$ fixed}} \, , 
\nonumber\\
\lim_{Q\to\infty} w_{f_1}(Q,Q) & = &  
 \pi^2 \, { 13 m^2 \over 3 Q^2} + \order\left( {1 \over Q^4 } 
\right) \, ,   \lbl{asympt_wf1}
\eea
and 
\bea
\lim_{Q_1\to\infty} w_{g_1}(M,Q_1,Q_2) & = &  
\pi^2 \, { M^2 [ Q_2^5 - Q_2 (Q_2^2 - 2 m^2) R_2^m] \over m^2 (Q_2^2 +
M_\pi^2) Q_1^3} + \order\left( {1 \over Q_1^5} \right) \, ,     
\quad{\mbox{ $Q_2$ fixed}}\,, 
\nonumber\\
\lim_{Q_2\to\infty} w_{g_1}(M,Q_1,Q_2) & = &  
\pi^2 \, {  M^2 [ - 18 m^2 Q_1^3 + 5 Q_1^5 + Q_1 (28 m^2 - 5 Q_1^2) R_1^m] 
\over 9 m^2 Q_2^5} \nonumber \\
&& + \, \order\left( {1 \over Q_2^7 } \right) ,\quad {\mbox{$Q_1$ fixed}},  
\nonumber\\
\lim_{Q\to\infty} w_{g_1}(M,Q,Q) & = & 
\pi^2 \, {  22 M^2 m^2 \over 3 Q^4}  + \order\left( {1 \over Q^5 }
\right) \, . 
\eea
Finally, the weight function $w_{g_2}(M,Q_1,Q_2)$ behaves as
follows:
\bea
\lim_{Q_1\to\infty} w_{g_2}(M,Q_1,Q_2) & = &  
\pi^2 \, {  [- Q_2^3 (Q_2^2 + m^2) + Q_2 (Q_2^2 - m^2) R_2^m] \over 3 m^2
Q_1^3}  + \order\left( {1 \over Q_1^5} \right) \, , 
\quad{\mbox{ $Q_2$ fixed}}\,, 
\nonumber\\
\lim_{Q_2\to\infty} w_{g_2}(M,Q_1,Q_2) & = &  
\pi^2 \, {  [ - Q_1^3 (Q_1^2 + 5 m^2) + Q_1 (Q_1^2 + 3 m^2) R_1^m] 
\over 3 m^2 Q_2^3} + \order\left( {1 \over Q_2^5 } \right) \, , 
\quad{\mbox{ $Q_1$ fixed}} \,, 
\nonumber\\
\lim_{Q\to\infty} w_{g_2}(M,Q,Q) & = & 
\pi^2 \, {  16 m^4 \over 9 Q^4} + \order\left( {1 \over Q^5 }
\right) \, .
\eea

Since $g_{M_{V_i}}^{(1)}(Q_1,Q_2) \equiv 0$ in the WZW model, the
corresponding integral $a_{\mu}^{\mbox{\tiny{LbyL;$\pi^0$}}(1)}$
involves only $w_{f_1}$ and diverges as ${\cal C} \ln^2\Lambda$ for some
ultraviolet cutoff $\Lambda$~\cite{melnikov01}. Varying the cutoff
$\Lambda$ in our numerical integration between 2 and 50 GeV, we obtain
${\cal C} \sim 0.025$. Since all form factors tend to the WZW model for $M_V
\to \infty$, the results for $a_{\mu}^{\mbox{\tiny{LbyL;$\pi^0$}}(1)}$ in
the different models should all scale as ${\cal C} \ln^2M_V$ for
large resonance masses, with the {\it same coefficient} ${\cal C}$ in
front of the log-squared term as in the WZW model. We have numerically
confirmed this observation for the VMD and LMD form factors. In fact,
for the WZW form factor, the region $Q_1 > Q_2$ in the integral
produces the $\ln^2 \Lambda$ term. Using the asymptotic expansion for
large $Q_1$ of $w_{f_1}(Q_1,Q_2)$ in Eq.~\rf{asympt_wf1}, one can
extract the coefficient of the leading-logarithm term from the integral,
with the result
\be \lbl{coeff_C} 
{\cal C} = 3 \left( { N_C \over 12 \pi} \right)^2 \left({m \over F_\pi}
\right)^2 \, . 
\ee
The corresponding value ${\cal C} = 0.0248$ (for $N_C=3$ and the
input values for $m$ and $F_\pi$ given in the next section) is in good
agreement with the numerical determination quoted above.  On the other
hand, $a_{\mu}^{\mbox{\tiny{LbyL;$\pi^0$}}}$ must vanish as the lepton
mass $m$ tends to zero. Numerically, we find that it is indeed the
case; see the next section.

Before we present our estimate for the pion-exchange contribution to
the muon anomalous magnetic moment in the next section, we would like
to stress again that the two-dimensional integral representation
\rf{api_two_dim}--\rf{api2} is valid for any form factor that can be written
as shown in Eq.~\rf{FF_f_g}. If in the future experimental data for
the off-shell form factor $\FF(-Q_1^2,-Q_2^2)$ become available, the
region below $1~\mbox{GeV}$ should also be measured with high
precision. Provided the data can be fitted with a representation of
the form factor belonging to the class we have discussed,
Eqs.~\rf{api_two_dim}--\rf{api2} can then be used to improve the
present estimates. In this sense, our integral representation is
similar to the familiar one that connects the vacuum polarization
contribution to $g-2$ and the experimental cross section $\sigma(e^+
e^- \to \mbox{hadrons})$~\cite{early_hadvacpol}.

\section{Numerical results}
\label{sec:numerics} 
\renewcommand{\theequation}{\arabic{section}.\arabic{equation}}
\setcounter{equation}{0}

The weight functions in Eqs.~\rf{wf1}--\rf{wg2M} are composed of
rational functions, square roots and logarithms. The combinations of
form factors from Eq.~\rf{hatf1_hatg2M} that enter in the integrals
are rational functions as well. It might therefore be possible to
perform one or even both integrations over $Q_1$ and $Q_2$
analytically for some parts of our expressions.\footnote{For the
analytical integration it might be convenient to extract the factor
$Q_1 Q_2$ from the weight functions and to rewrite the integrals in
terms of $Q_1^2$ and $Q_2^2$.}  On the other hand, the numerical
evaluation of the two-dimensional integral representation for
$a_{\mu}^{\mbox{\tiny{LbyL;$\pi^0$}}}$ in Eqs.~\rf{api1} and \rf{api2}
poses no real problems. The corresponding results for the different
form factors discussed earlier can be found in
Table~\ref{tab:api_models}. Apart from the masses of the muon $m =
m_\mu = 105.66~\mbox{MeV}$ and the pion $M_{\pi^0} =
134.98~\mbox{MeV}$, we have used the following input values: $\alpha =
1/ 137.03599976, F_\pi = 92.4~\mbox{MeV}, M_{V_1} = M_\rho =
769~\mbox{MeV},$ and $M_{V_2} = M_{\rho^\prime} =
1465~\mbox{MeV}$~\cite{PDG00}. The errors from the numerical
integration are much smaller than the last digits given in the table.

\begin{table}[h]
\caption{Results for the terms $a_{\mu}^{\mbox{\tiny{LbyL;$\pi^0$}}(1)}$, 
$a_{\mu}^{\mbox{\tiny{LbyL;$\pi^0$}}(2)}$
and for the pion-exchange contribution to the anomalous magnetic
moment $a_{\mu}^{\mbox{\tiny{LbyL;$\pi^0$}}}$ 
according to Eq.~{\protect\rf{api_two_dim}} for
the different form factors considered. In the WZW model we used a cutoff of
$1~\mbox{GeV}$ in the first contribution, whereas the second term is
ultraviolet finite. In the LMD+V {\it Ansatz} we used $h_1 =
0~\mbox{GeV}^2$ and $h_5 = 6.93~\mbox{GeV}^4$.}
\begin{center}
\renewcommand{\arraystretch}{1.1}
\begin{tabular}{|l|r@{.}l|r@{.}l|r@{.}l|}
\hline
Form factor &
\multicolumn{2}{|c|}{{$a_{\mu}^{\mbox{\tiny{LbyL;$\pi^0$}}(1)}$}}   
      & \multicolumn{2}{|c|}{{$a_{\mu}^{\mbox{\tiny{LbyL;$\pi^0$}}(2)}$}}  
      & \multicolumn{2}{|c|}{{$a_{\mu}^{\mbox{\tiny{LbyL;$\pi^0$}}}
\times 10^{10}$}}   
\\ 
\hline  
WZW 	& \hspace*{0.25cm} 0 & 095  	& \hspace*{0.25cm}0 & 0020 &
\hspace*{0.65cm}  12 & 2 \\  
VMD	& 0 & 044 	& 0 & 0013	& 5 & 6 \\ 
LMD 	& 0 & 057 	& 0 & 0014 	& 7 & 3 \\
LMD+V ($h_2 = - 10~\mbox{GeV}^2$)
        & 0 & 049	& 0 & 0013 	& 6 & 3 \\
LMD+V ($h_2 = 0~\mbox{GeV}^2$)
        & 0 & 045	& 0 & 0013 	& 5 & 8 \\
LMD+V ($h_2 = 10~\mbox{GeV}^2$)
        & 0 & 041	& 0 & 0013 	& 5 & 3 \\
\hline
\end{tabular}
\label{tab:api_models}  
\end{center}
\end{table}

As already announced in the Introduction we obtain a different sign
for $a_{\mu}^{\mbox{\tiny{LbyL;$\pi^0$}}}$ as compared to the latest
two calculations~\cite{HKS_95_96,HK_98,BPP}. This is immediately
visible from the plots of the weight functions $w_{f_1}$ and $w_{g_1}$
which are positive and which are multiplied by positive functions
$f^{(1)}$ and $g_{M_V}^{(1)}$, at least for the WZW, VMD, and LMD 
form factors. We shall further discuss this important point in the next
section. On the other hand the contribution
$a_{\mu}^{\mbox{\tiny{LbyL;$\pi^0$}}(2)}$ is highly suppressed and
very stable with respect to the various form factors (with the obvious
exception of the WZW case). As noted earlier, the corresponding
integral converges even for a constant form factor as in the WZW
model. As far as the numerics is concerned, the absolute value for
$a_{\mu}^{\mbox{\tiny{LbyL;$\pi^0$}}}$ for the VMD from factor is
identical to the results quoted in
Refs.~\cite{KNO_85,HKS_95_96,HK_98,BPP}, whereas the one for the LMD
form factor agrees, up to the sign, with the result for the form
factor number 4 given in Ref.~\cite{Bijnens_Persson}.

Since the results for the cases of the LMD+V and VMD form factors are
very similar, one might conclude that imposing the asymptotic $1/Q^2$
behavior of $\FF(-Q^2,0)$ for large spacelike momenta is more
important than reproducing the QCD short-distance behavior for $Q_1^2
\to \infty$ and $Q_2^2 \to \infty$, as does the LMD form
factor. However, the slightly higher result for
$a_{\mu}^{\mbox{\tiny{LbyL;$\pi^0$}}}$ in the case of the LMD form
factor might also be due to the fact that in this instance the slope
of the form factor at the origin, as obtained, for instance, by the
CELLO Collaboration~\cite{CELLO}, is not well reproduced, in contrast
to the LMD+V and VMD cases (see the discussion in
Ref.~\cite{paper_VAP}). As noted in connection with the plots of the
weight functions, it is mainly the region $Q_i \lapprox 1~\mbox{GeV}$
that contributes in the integrals.

For illustration, we decompose, in the cases of the LMD and LMD+V form
factors, the total numerical result of Table~\ref{tab:api_models}
according to the integrals corresponding to the different weight
functions appearing in the sums in Eqs.~\rf{api1} and \rf{api2},
\bea
&&
\left. 
\begin{array}{rclcl}
a_{\mu}^{\mbox{\tiny{LbyL;$\pi^0$}}(1)} & = & 0.015 + 0.042 + 0 
& = & 0.057 \\ 
a_{\mu}^{\mbox{\tiny{LbyL;$\pi^0$}}(2)} & = & 0.0016 - 0.0002 + 0 
& = & 0.0014 \\ 
\end{array}
\right\} \mbox{LMD}\, ,  \\
&&
\left. 
\begin{array}{rclcl}
a_{\mu}^{\mbox{\tiny{LbyL;$\pi^0$}}(1)} & = & 0.0026 + 0.0448 - 0.0026
& = & 0.045 \\ 
a_{\mu}^{\mbox{\tiny{LbyL;$\pi^0$}}(2)} & = & 0.0015 - 0.0002 - 1
\times 10^{-6} 
& = & 0.0013 
\end{array}
\right\} \mbox{LMD+V}~(h_2 = 0) \, . 
\eea
One observes that in both cases the term involving the weight function 
$w_{g_1}(M_{V_1},Q_1,Q_2)$ gives the main contribution to the final
result. 

For a fixed value of $h_2$ in the LMD+V form factor, our results are
rather stable under the variation of the other parameters. For
instance, varying $M_{V_1}$, $M_{V_2}$, and $h_5$ by $\pm
20~\mbox{MeV}$, $\pm 25~\mbox{MeV}$, and $\pm 0.5~\mbox{GeV}^4$,
respectively, the result for $a_{\mu}^{\mbox{\tiny{LbyL;$\pi^0$}}}$
changes by $\pm 0.2 \times 10^{-10}$. On the other hand, if all other
parameters are kept fixed, our result depends almost linearly on
$h_2$, at least for $|h_2| < 20~\mbox{GeV}^2$. In this range
$a_{\mu}^{\mbox{\tiny{LbyL;$\pi^0$}}}$ changes by $\pm 0.9 \times
10^{-10}$ from the central value for $h_2 = 0$. As noted earlier, the
experimental data for $\pi^0 \to e^+ e^-$~\cite{alavi99,PDG00} seem to
favor slightly positive values for $h_2$, although the bounds are
rather loose. A better experimental determination of this decay rate
would therefore be highly welcome.

Thus, using the LMD+V form factor, our estimate for the muon anomalous
magnetic moment from the pion-pole contribution in hadronic
light-by-light scattering reads
\be 
a_{\mu}^{\mbox{\tiny{LbyL;$\pi^0$}}} = + 5.8~(1.0) \times 10^{-10}
\, , 
\ee
where the error includes the variation of the parameters and the
intrinsic model dependence.

As far as the contribution to $a_\mu$ from the exchange of the $\eta$
or of the $\eta^\prime$ is concerned, a detailed short-distance analysis
of the corresponding form factor in large-$N_C$ QCD for nonzero quark
masses as done for the pion is beyond the scope of the present
work. Furthermore, as noted above, the integrals in Eqs.~\rf{api1} and
\rf{api2} do not seem to be very sensitive to the correct asymptotic
behavior for large momenta. It is more important to have a good
description at small and intermediate energies, e.g., by reproducing
the slope of the form factor ${\cal
F}_{\mbox{\tiny{PS}}\gamma^*\gamma^*}(-Q^2,0)$, $\mbox{PS} =
\eta,\eta^\prime$, at the origin (cf.\ the similar results obtained
for the VMD and the LMD+V form factors). The CLEO
Collaboration~\cite{CLEO} has made a fit of the form factors ${\cal
F}_{\eta\gamma^*\gamma^*}(-Q^2,0)$ and ${\cal F}_{\eta^\prime
\gamma^*\gamma^*}(-Q^2,0)$, normalized to the corresponding 
experimental width
$\Gamma(\mbox{PS} \to \gamma \gamma)$, using a VMD {\it Ansatz} with an
adjustable vector meson mass $\Lambda_{\mbox{{\tiny PS}}}$. Taking
their values $\Lambda_\eta = 774 \pm 29~\mbox{MeV}$ or
$\Lambda_{\eta^\prime} = 859 \pm 28~\mbox{MeV}$ as the vector meson mass
$M_V$ in the expression of the VMD form factor, we obtain from
Eq.~\rf{api_two_dim}
\bea
a_{\mu}^{\mbox{\tiny{LbyL;$\eta$}}}\vert_{VMD} 
& = & + 1.3~(0.1) \times 10^{-10} \, ,
\nonumber \\ 
a_{\mu}^{\mbox{\tiny{LbyL;$\eta^\prime$}}}\vert_{VMD} 
& = & + 1.2~(0.1) \times 
10^{-10} \, . \lbl{a_mu_eta_etaprime} 
\eea
The error reflects only the corresponding variation in
$\Lambda_{\mbox{\tiny{PS}}}$. Again, apart from the {\it sign}, these values
are comparable with the numbers quoted in
Refs.~\cite{HKS_95_96,HK_98,BPP,Bijnens_Persson}. We do not expect the
results with the LMD+V form factor to differ much from the numbers
quoted above. Thus, adding up all contributions from pseudoscalar
exchange, we obtain
\be
a_{\mu}^{\mbox{\tiny{LbyL;PS}}} \equiv a_{\mu}^{\mbox{\tiny{LbyL;$\pi^0$}}}
+ a_{\mu}^{\mbox{\tiny{LbyL;$\eta$}}}\vert_{VMD} +
a_{\mu}^{\mbox{\tiny{LbyL;$\eta^\prime$}}}\vert_{VMD} = + 8.3~(1.2) \times
10^{-10} \, . 
\ee

Finally, for illustration and as a check, the result for the pion
exchange contribution to the anomalous magnetic moment of the {\it
electron} reads 
\bea
a_{e}^{\mbox{\tiny{LbyL;$\pi^0$}}(1)} & = & 2.0 \times 10^{-6}
\nonumber \, , \\ 
a_{e}^{\mbox{\tiny{LbyL;$\pi^0$}}(2)} & = & 2.0 \times 10^{-6}
\nonumber \, ,  \\ 
a_{e}^{\mbox{\tiny{LbyL;$\pi^0$}}}    & = & 5.1 \times 10^{-14} \, ,
\eea
for the LMD+V form factor (with $h_2 = 0$). Despite the various
factors $1/m^2$ in the weight functions, the final result is much
smaller than for the muon, as it should be.  The results for the other
form factors are of similar size (LMD:
$a_{e}^{\mbox{\tiny{LbyL;$\pi^0$}}} = 7.7 \times 10^{-14}$, VMD:
$a_{e}^{\mbox{\tiny{LbyL;$\pi^0$}}} = 2.6 \times 10^{-14}$), except for
the WZW case. Note that for the electron the second contribution
$a_{e}^{\mbox{\tiny{LbyL;$\pi^0$}}(2)}$ is in general not smaller than
the first one, in contrast to the muon case.

\section{Discussion and conclusions} 
\label{sec:conclusions} 
\renewcommand{\theequation}{\arabic{section}.\arabic{equation}}
\setcounter{equation}{0}

In this article, we have evaluated the correction to $a_\mu$ induced
by the pion-pole contribution to hadronic light-by-light scattering.
From the methodological point of view, the present approach differs
from previous ones in two important points. First, we use a
representation of the form factor $\FF$ which incorporates
short-distance properties of QCD within an approximation to the
large-$N_C$ limit involving a finite number of resonances. It is
important to stress that this approximation can in principle be
improved by considering additional vector meson states, if additional
information, subleading short-distance corrections, phenomenological
and/or experimental input, is provided. Second, for the corresponding
class of form factors, we have performed the angular integration in
the two-loop expression in an analytical way, using the method of
Gegenbauer polynomials. This leads to an expression for 
$a_{\mu}^{\mbox{\tiny{LbyL;$\pi^0$}}}$ in terms of weighted
two-dimensional integrals over the moduli of the Euclidean loop
momenta.  These integrals were then evaluated numerically, and the
results displayed in Table 2.

Whereas the size of $a_{\mu}^{\mbox{\tiny{LbyL;$\pi^0$}}}$ obtained in
this way is quite comparable to the existing determinations, the sign
comes out opposite.  In view of the consequences implied by this
result, it is clear that this discrepancy needs to be discussed. We
first notice that our result is not an artifact due to the
representation of the form factor we use.  Indeed, repeating the same
analysis with the VMD form factor, we exactly reproduce the result
quoted by previous authors, but again {\it with the opposite global
sign}. Now, one might observe that the VMD form factor contributes
only to the second term in Eq. \rf{api1}, involving the positive
definite (see Fig.~\ref{fig:fig3}) weight function
$w_{g_1}(M_V,Q_1,Q_2)$, so that our result in this case could be
ascribed to the fact that we obtain the wrong sign for that
function. In addition to the various checks that we have performed, we
shall however put forward a few arguments that support the correctness
of our result.  In the case of the constant form factor $\FF^{WZW}$, only
the first term in Eq. \rf{api1}, involving the positive definite
weight function $w_{f_1}(Q_1,Q_2)$, contributes. This integral is
divergent, and behaves as
\cite{melnikov01,KNPdR01}
\be
\lim_{\Lambda\to\infty} \int_0^\Lambda dQ_1 \int_0^\Lambda dQ_2
\,\left(\frac{N_C}{12\pi^2F_\pi}\right)^2
w_{f_1}(Q_1,Q_2) \,=\, {\cal C}\ln^2 \Lambda\,+\,\cdots 
\lbl{div} 
\ee
for a large ultraviolet cutoff $\Lambda$ (the ellipsis stands for
subleading $\ln\Lambda$ divergences, and for finite terms).  The
coefficient ${\cal C}$ of this log-squared dependence can be computed
analytically in a way \cite{KNPdR01} that is largely independent of
the methods used here. It is {\it positive} and agrees with the value
we obtain from a numerical analysis of our formulas and from the
asymptotic behavior for large momenta of the weight function
$w_{f_1}(Q_1,Q_2)$; see Eq.~\rf{coeff_C}. Next, as $M_V$ becomes very
large, the VMD form factor approaches the WZW form factor and
therefore should reproduce the previous result,
\be
\lim_{M_V\to\infty} \int_0^\infty dQ_1 \int_0^\infty dQ_2
\,w_{g_1}(M_{V},Q_1,Q_2) \
g_{M_{V}}^{(1)}(Q_1^2, Q_2^2)\,=\,{\cal C}\ln^2 M_V\,+\,\cdots
\ee
{\it with the same constant} ${\cal C}$ as in Eq.~\rf{div}.  We have checked
numerically that this indeed happens with Eq. \rf{api1} and the
expression of $w_{g_1}(M_V,Q_1,Q_2)$ given in
Eq. \rf{hatf1_hatg2M}. Thus, the signs of the contributions involving
$w_{f_1}$ {\it and} $w_{g_1}$ must be correct. 

Finally, we mention that with the convention for the external momenta
used in the third article quoted under Ref. \cite{BPP}, the right-hand
side of Eq. (2.7) in that paper, which corresponds to our
Eq. \rf{F2trace}, should come with a global minus sign.  We also
stress that our Eq. \rf{F2trace} agrees with Eq. (2.9) of \cite{ABDK}.
We have not been able to find a similar point of disagreement with
Refs. \cite{HKS_95_96,HK_98}. However, the lack of analytical
expressions in intermediate steps precludes a more detailed
comparison. Clearly, this matter needs to be investigated further in
the future.\footnote{{\it Note added:} After the submission of this
work, both groups~\cite{HK_01_BPP_01} found the sign error in their
calculations in Refs.~\cite{HKS_95_96,HK_98,BPP}. Furthermore,
in the case of the VMD form factor, the analytical result was derived
in Ref.~\cite{BCM_01}, in the form of a double series in $(M_{\pi^0}^2
- m^2) / m^2$ and $m^2 / M_V^2$. The numerical value agrees exactly
with our result for this form factor. Independently, Bardeen and de
Gouvea~\cite{Bardeen_deGouvea} have reproduced the coefficient of the
log-squared term from Eq.~\rf{coeff_C} by an analytical calculation.}

Concluding this study, we find that the pseudoscalar-pole contribution
to the hadronic light-by-light scattering amounts to
\be \lbl{final_result}
a_{\mu}^{\mbox{\tiny{LbyL;PS}}}\,=\,+\,8.3~(1.2) \times 10^{-10}  \,, 
\ee
where we have used the LMD+V representation \rf{FF_LMD+V} of the form
factor $\FF$ and the VMD form factor for the $\eta$ and $\eta^\prime$,
both supplemented with experimental information. The error given in
Eq.~\rf{final_result} includes our estimate of the intrinsic model
dependence and the variation of the model parameters. If we assume
that the other contributions to the hadronic light-by-light scattering
remain unchanged, our result implies that the difference between
theory and experiment reduces to about $25~(16)\times 10^{-10}$, in
the ``least favorable'' case, i.e., the vacuum polarization
analysis of Ref. \cite{davier}, which has the smallest value and the
smallest error as compared to other recent evaluations
\cite{narison01,jegerlehner01,yndurain01,cvetic01,prades01}. We intend to
use the same approach to study the full four-point function
$\Pi_{\mu\nu\lambda\rho}(q_1,q_2,q_3)$ in order to obtain an
evaluation of the complete contribution
$a_{\mu}^{\mbox{\tiny{LbyL;had}}}$.  We also plan to investigate in
more detail the $\eta$ and $\eta^\prime$ pole contributions, although
we do not expect the numerical values to change significantly as
compared to the ones obtained in Eq.~\rf{a_mu_eta_etaprime} with the
VMD form factor.

\section*{Acknowledgments} 

We are grateful to M.~Perrottet and E.~de Rafael for many stimulating
discussions, encouragement, and for sharing their insight on various
aspects of the muon $g-2$.  We further thank T. Becher and V. Cuplov
for their kind assistance, as well as J. Gasser for discussions.  This
work was supported in part by Schweizerischer Nationalfonds and by
TMR, EC Contract No.\ ERBFMRX-CT980169 (EURODA$\Phi$NE).

\newpage
\section*{Appendix}
\renewcommand{\theequation}{A\arabic{equation}}
\setcounter{equation}{0}

\indent

\noindent
This Appendix lists the properties of the form factor
$\FF(q_1^2,q_2^2)$ that are reproduced by the LMD and LMD+V
representations.

At the low-energy end, the form factor is normalized by the 
$\pi^0\to\gamma\gamma$ amplitude
\be
e^2 \FF(0,0)\,=\,{\cal A}(\pi^0\to\gamma\gamma)\,.
\ee
In the chiral limit $m_q\to 0$, $q=u,d,s$, this amplitude is fixed by the 
WZW anomaly to read
\be
{\cal A}^{(0)}(\pi^0\to\gamma\gamma)\,=\,-\,\frac{ e^2N_C}{12\pi^2F_0}\,.
\ee
For massive light quarks, this expression receives corrections. In
particular, the pion decay constant in the chiral limit, $F_0$, is
replaced by its physical counterpart $F_\pi = F_0[1+{\cal O}(m_q)]$,
\be
{\cal
A}(\pi^0\to\gamma\gamma)\,=\,-\,\frac{ e^2N_C}{12\pi^2F_\pi}\,[1+{\cal
O}(m_q)]\,. 
\ee
It turns out that the additional ${\cal O}(m_q)$ corrections in this
relation are numerically small
\cite{Pseudoscalar_decays}, so that one may drop them to
a good approximation.

The behavior of $\FF$ at short distances was recently studied in
Ref. \cite{paper_VAP} (see also the references therein) in the chiral
limit.  It turns out that the properties relevant for our present
purposes still hold for massive light quarks, again upon replacing
$F_0$ by $F_\pi$,
\be
\lim_{\lambda\to \infty}\,\FF(\lambda^2 q^2, (p-\lambda q)^2)\,=\,
\frac{2}{3}\,\frac{F_\pi}{q^2}\,\bigg\{ \frac{1}{\lambda^2}\,+\,
\frac{1}{\lambda^3}\,\frac{q\cdot p}{q^2}\,+\,{\cal
O}\left({1 \over \lambda^4 }\right)\bigg\}\,. 
\lbl{OPE_FF}
\ee
This expression holds up to ${\cal O}(\alpha_s)$ corrections. Note, 
however, that the Wilson coefficients corresponding to the two first
terms of the short-distance expansion shown here are free of anomalous
dimensions.

In the large-$N_C$ limit of QCD, the singularities of $\FF$ are 
restricted to single poles in each channel,
\be
\FF(q_1^2,q_2^2)\big\vert_{N_C\to\infty}\,=\,
\sum_{ij}\frac{c_{ij}(q_1^2,q_2^2)}{(q_1^2-M_{V_i}^2)(q_2^2-M_{V_j}^2)}
\,,
\lbl{large_Nc_FF}
\ee
where the sum runs over the infinite tower of zero-width $J^{PC} =
1^{--}$ vector resonances of large-$N_C$ QCD. Beyond the fact that
they must reproduce the short-distance behavior \rf{OPE_FF} and be
free of singularities, the functions $c_{ij}(q_1^2,q_2^2)$ are not
further restricted by the (known) properties of large-$N_C$ QCD. The
representations $\FF^{LMD}$ and $\FF^{LMD+V}$ are obtained by
truncation of the infinite sum \rf{large_Nc_FF} to one, respectively
two, vector resonances per channel. As noted in the text, as $M_V \to
\infty$, the form factor $\FF^{LMD}$ reduces to a constant given by the 
Wess-Zumino-Witten term. In the case of $\FF^{LMD+V}$ one has to proceed in
two steps. First the heavier resonance $V_2$ is decoupled, 
\be
\lim_{M^2_{V_2} \to \infty} \FF^{LMD+V}(q_1^2,q_2^2) =
\FF^{LMD}(q_1^2,q_2^2) \, , 
\ee
with 
\be
\lim_{M^2_{V_2} \to \infty} {h_7 \over M_{V_2}^4} = - c_V, \quad 
\lim_{M^2_{V_2} \to \infty} {h_5 \over M_{V_2}^4} = 1 \, , \quad 
\lim_{M^2_{V_2} \to \infty} {h_i \over M_{V_2}^4} = 0, \ \ i=1,2\, . 
\ee
Then one lets $M_{V_1} \to \infty$ as before.


\begin{thebibliography}{99} 


\bibitem{BNL1}
R.~M.~Carey {\it et al.},
Phys.\ Rev.\ Lett.\  {\bf 82}, 1632 (1999);
Muon $g-2$ Collaboration, H.~N.~Brown {\it et al.},
Phys.\ Rev.\ D {\bf 62}, 091101 (2000).

\bibitem{BNL2}
Muon $g-2$ Collaboration, H.~N.~Brown {\it et al.}, 
Phys.\ Rev.\ Lett.\  {\bf 86}, 2227 (2001).

\bibitem{cern77}
J. Bailey {\it et al.}, Phys. Lett. {\bf 68B}, 191 (1977).


\bibitem{KNO90}
T.\ Kinoshita, B.\ Ni\v zi\'c, and Y.\ Okamoto, Phys.\ Rev.\ D 
{\bf 41}, 593 (1990). 

\bibitem{KinMar}
T. Kinoshita and W.~J. Marciano, 
in: {\it Quantum Electrodynamics}, edited by T.\ Kinoshita (Advanced
Series on Directions in High Energy Physics, Vol.\ 7) (World
Scientific, Singapore, 1990), p.~419. 

\bibitem{spires}
A list of references can be obtained from the URL\\   
http://www.slac.stanford.edu/spires/find/hep?c=PRLTA,86,2227.

\bibitem{PPdR95}
S. Peris, M. Perrottet, and E. de Rafael, Phys. Lett. B {\bf 355}, 523
(1995). 

\bibitem{CzKM95}
A. Czarnecki, B. Krause, and W.J. Marciano, Phys. Rev. D {\bf 52},
R2619 (1995).  

\bibitem{KPPdR01}
M. Knecht, S. Peris, M. Perrottet, and E. de Rafael (in preparation). 

\bibitem{early_hadvacpol}
C. Bouchiat and L. Michel, J.\ Phys.\ Radium {\bf 22}, 121 (1961); 
%
L. Durand III, Phys.\ Rev.\ {\bf 128}, 441 (1962); {\bf 129},
2835(E) (1963); 
%
M. Gourdin and E. de Rafael, Nucl. Phys. {\bf B10}, 667 (1969).

\bibitem{davier}
M. Davier and A. H\"ocker, Phys. Lett. B {\bf 435}, 427 (1998).

\bibitem{narison01}
S.~Narison,
Phys.\ Lett.\ B {\bf 513}, 53 (2001); {\bf 526}, 414(E) (2002). 

\bibitem{jegerlehner01}
F. Jegerlehner, hep-ph/0104304.

\bibitem{yndurain01}
J.~F. de Troc\'oniz and F.~J. Yndur\'ain, Phys.\ Rev.\ D (to be
published); hep-ph/0106025.

\bibitem{cvetic01}
G.~Cveti\v c, T.~Lee, and I.~Schmidt,
Phys.\ Lett.\ B {\bf 520}, 222 (2001). 

\bibitem{prades01}
J. Prades, in Proceedings of the International Conference on {\it CP}
Violation (KAON2001), Pisa, Italy, 
2001 (to be published); hep-ph/0108192.

\bibitem{cirigliano}
V.~Cirigliano, G.~Ecker, and H.~Neufeld,
Phys.\ Lett.\ B {\bf 513}, 361 (2001). 

\bibitem{calmet76}
J. Calmet, S. Narison, M. Perrottet, and E. de Rafael, Phys. Lett. {\bf
61B}, 283 (1976); Rev. Mod. Phys. {\bf 49}, 21 (1977).

\bibitem{KNO_85}
T.\ Kinoshita, B.\ Ni\v zi\'c, and Y.\ Okamoto,  Phys. Rev. Lett. {\bf
52}, 717 (1984); 
Phys.\ Rev.\ D {\bf 31}, 2108 (1985);
Y.\ Okamoto, Ph.D. thesis, Cornell University, Ithaca, NY, 1984.


\bibitem{HKS_95_96}
M.~Hayakawa, T.~Kinoshita, and A.~I.~Sanda,
Phys.\ Rev.\ Lett.\  {\bf 75}, 790 (1995); 
%
%
Phys.\ Rev.\ D {\bf 54}, 3137 (1996). 

\bibitem{HK_98}
M.\ Hayakawa and T.\ Kinoshita, 
Phys.\ Rev.\ D {\bf 57}, 465 (1998). 

\bibitem{BPP}
J.\ Bijnens, E.\ Pallante, and J.\ Prades, 
Phys.\ Rev.\ Lett.\ {\bf 75}, 1447 (1995); {\bf 75}, 3781(E)
(1995);  
%
%
Nucl.\ Phys.\ {\bf B474}, 379 (1996). 

\bibitem{bartos01}
E. Barto{\v s} {\it et al.}, hep-ph/0106084~v2.

\bibitem{tHooft74}
G. 't Hooft, Nucl. Phys. {\bf B72}, 461 (1974); {\bf B75}, 461 (1974).

\bibitem{witten79}
E. Witten, Nucl. Phys. {\bf B160}, 57 (1979).

\bibitem{derafael01}
E. de Rafael, in Proceedings of the International Workshop on QCD:
Theory and Experiment (QCD @ Work), Martina Franca, 
Italy, 
2001 (to be published); hep-ph/0110195.

\bibitem{wilson69}
K.~G. Wilson, Phys. Rev. {\bf 179}, 1499 (1969).

\bibitem{SVZ}
M.~A. Shifman, A.~I. Vainshtein, and V.~I. Zakharov, 
Nucl. Phys. {\bf B147}, 385 (1979); {\bf B147}, 447 (1979).

\bibitem{paper_VAP}
M.\ Knecht and A.\ Nyf\/feler, 
Eur. Phys. J. C {\bf 21}, 659 (2001).

\bibitem{derafael94}
E. de Rafael, Phys. Lett. B {\bf 322}, 239 (1994).

\bibitem{WZW}
J.\ Wess and B.\ Zumino, Phys.\ Lett.\  {\bf 37B}, 95 (1971); 
E.\ Witten, Nucl.\ Phys.\ {\bf B223}, 422 (1983). 

\bibitem{ABDK}
J.\ Aldins, S.~J.\ Brodsky, A.~J.\ Dufner, and T.\ Kinoshita, 
Phys.\ Rev.\ D {\bf 1}, 2378 (1970). 

\bibitem{BS67}
S.~J. Brodsky and J.~D. Sullivan, Phys. Rev. {\bf 156}, 1644 (1967).


\bibitem{BR}
R.\ Barbieri and E.\ Remiddi, 
Nucl.\ Phys.\ {\bf B90}, 233
(1975). 

\bibitem{RRL}
R.~Z.\ Roskies, E.\ Remiddi, and M.~J.\ Levine, 
in {\it Quantum Electrodynamics}~\cite{KinMar}, p.~162. 

\bibitem{hearn}
A.~C. Hearn, {\small{REDUCE}} User's Manual Version 3.5, RAND
Publication CP78, 1993. 

\bibitem{CLEO}
J. Gronberg {\it et al.}, Phys. Rev. D {\bf 57}, 33 (1998). 

\bibitem{ABJ}
S.~L.\ Adler, Phys. Rev. {\bf 177}, 2426 (1969); 
J.~S.\ Bell and R.\ Jackiw, Nuovo Cimento A {\bf 60}, 47 (1969).  

\bibitem{Bijnens_Persson}
J.\ Bijnens and F.\ Persson, 
hep-ph/0106130. 

\bibitem{Brodsky_Lepage}
G.~P. Lepage and S.~J. Brodsky, Phys. Lett. {\bf 87B}, 359 (1979); 
Phys. Rev. D {\bf 22}, 2157 (1980); S.~J. Brodsky and G.~P. Lepage,
{\it ibid.} {\bf 24}, 1808 (1981). 

\bibitem{Pi_ll}
M.\ Knecht, S.\ Peris, M.\ Perrottet, and E.\ de Rafael, Phys.\ Rev.\
Lett.\ {\bf 83}, 5230 (1999). 


\bibitem{alavi99}
KTeV Collaboration, A.~Alavi-Harati {\it et al.},
Phys.\ Rev.\ Lett.\  {\bf 83}, 922 (1999).

\bibitem{PDG00}
Particle Data Group, D.~E.~Groom {\it et al.},
Eur.\ Phys.\ J.\ C {\bf 15}, 1 (2000).


\bibitem{early_Gegenbauer}
J.~L.\ Rosner, 
Ann.\ Phys.\ (N.Y.) {\bf 44}, 11 (1967). Implicitly, the method of
Gegenbauer polynomials was already used in the following papers: M.\
Baker, K.\ Johnson, and R.\ Willey, 
Phys.\ Rev. {\bf 136}, B1111 (1964); 
{\bf 163}, 1699 (1967).


\bibitem{LRR}
M.~J.\ Levine and R.\ Roskies, 
Phys.\ Rev.\ D {\bf 9},
421 (1974); M.~J.\ Levine, E.\ Remiddi, and R.\ Roskies, 
{\it ibid.} {\bf 20}, 2068 (1979).

\bibitem{master} 
D. Kreimer, Phys.\ Lett.\ B {\bf 273}, 277 (1991);
G.\ Weiglein, R. Scharf, and M.\ B\"ohm, Nucl.\ Phys.\ {\bf B416}, 606 (1994);
S.\ Bauberger, M.\ B\"ohm, G.\ Weiglein, F.~A.\ Berends, and M.\ Buza,
Nucl.\ Phys.\ B (Proc.\ Suppl.) {\bf 37B}, 95 (1994);
S.\ Bauberger, F.~A.\ Berends, M.\ B\"ohm, and M.\ Buza, 
Nucl.\ Phys.\ {\bf B434}, 383 (1995);
S.\ Bauberger and M.\ B\"ohm, 
{\it ibid.} {\bf B445}, 25 (1995); 
O.~V. Tarasov, {\it ibid.} {\bf B502}, 455 (1997).


\bibitem{MOS}
W.\ Magnus, F.\ Oberhettinger, and R.P.\ Soni, {\it Formulas and
Theorems for the Special Functions of Mathematical Physics} 
(Springer, New York, 1966), p.~218. 

\bibitem{AS}
{\it Handbook of Mathematical Functions}, edited by M.\ Abramowitz and
I.~A.\ Stegun (Dover Publications, New York, 1970), p.~771.  

\bibitem{melnikov01}
K. Melnikov, 
Int.\ J.\ Mod.\ Phys.\ A {\bf 16}, 4591 (2001). 

\bibitem{CELLO}
H.-J. Behrend {\it et al.}, Z. Phys. C {\bf 49}, 401 (1991). 

\bibitem{KNPdR01}
M. Knecht, A. Nyf\/feler, M. Perrottet, and E. de Rafael, Phys.\ Rev.\
Lett.\ {\bf 88}, 071802 (2002).  

\bibitem{HK_01_BPP_01}
M.~Hayakawa and T.~Kinoshita (private communication); 
hep-ph/0112102; 
%
J.~Bijnens, E.~Pallante, and J.~Prades, hep-ph/0112255. 

\bibitem{BCM_01}
I.~Blokland, A.~Czarnecki, and K.~Melnikov, Phys.\ Rev.\ Lett.\ {\bf
88}, 071803 (2002). 

\bibitem{Bardeen_deGouvea}
W.\ Bardeen and A.\ de Gouvea (private communication, December 3, 2001). 

\bibitem{Pseudoscalar_decays}
J.~Bijnens, A.~Bramon, and F.~Cornet,
Phys.\ Rev.\ Lett.\  {\bf 61}, 1453 (1988); 
%
B.~Moussallam,
Phys.\ Rev.\ D {\bf 51}, 4939 (1995); 
%
L.~Ametller, J.~Kambor, M.~Knecht, and P.~Talavera,
{\it ibid.} {\bf 60}, 094003 (1999).

\end{thebibliography}
\end{document}